\pdfoutput=1

\documentclass[article]{mesa2}
\usepackage{times,mathptm}
\usepackage{amsmath,amsfonts,amssymb}
\usepackage{graphicx,graphics,epsfig}
\usepackage{fancyhdr,fancybox}
\usepackage{verbatim}
\usepackage{color}
\usepackage{subfig}
\jvol{3}
\jnum{2}
\jyear{2012}

\numberwithin{equation}{section}
\numberwithin{theorem}{section}
\numberwithin{definition}{section}
\numberwithin{lemma}{section}
\numberwithin{corollary}{section}
\numberwithin{proposition}{section}
\numberwithin{example}{section}
\numberwithin{remark}{section}

\begin{document}
\label{pageinit}

\date{}

\title{A new dynamical indicator for chaos detection in galactic Hamiltonian systems}

\author{Euaggelos E. Zotos$^{\star}$}

\maketitle

\noindent Department of Physics, Section of Astrophysics, Astronomy and Mechanics, \\
Aristotle University of Thessaloniki, GR-541 24 Thessaloniki, Greece.\\

\noindent $^\star$ \emph{Corresponding author}: evzotos@astro.auth.gr

\markboth{Euaggelos E. Zotos}{A new dynamical indicator for chaos detection in galactic Hamiltonian systems}


\begin{abstract}
A new dynamical parameter, the $f$-indicator, is introduced and used in order to distinguish between regular and chaotic motion in galactic Hamiltonian systems. Two kinds of galactic potentials are used: (i) a global potential, which describes the whole galaxy and (ii) a local potential, which is made up of perturbed harmonic oscillators and describes motion near an equilibrium point. The new indicator is based on the energies of the separable system along the $x, y$ and $z$ axis. Comparison between the outcomes obtained using the new dynamical parameter and other methods, such as the maximum Lyapunov Characteristic Exponent (L.C.E), or the $S(c)$ dynamical spectrum, shows that the new dynamical indicator gives fast and reliable results concerning the regular or chaotic character of the orbits. The new indicator was tested in several Hamiltonian systems of two (2D) degrees and three (3D) degrees of freedom.
\end{abstract}
%%%%%%%%%%%%%%%%%%%%%%%%%%%%%%%%%%%%%%%%%%%%%%%%%%%%%%%%%%%%%%%%%%%%%%%%%%%%%%%%%%%%%%%%%%%%%%%%%%%%

\section{Introduction}

Knowing whether the orbits of a dynamical system are ordered or chaotic, is fundamental for the understanding of the behavior of the Hamiltonian system. In the dissipative case, the distinction is easily made, as both types of motion (regular or chaotic) are attracting. In conservative systems, however, distinguishing between order and chaos is often a delicate issue (e.g., when the chaotic or regular regions are small), especially in Hamiltonian systems with many degrees of freedom, where one cannot easily visualize the structure of the phase space. For this reason, it is of great importance to have quantities and indicators, in order to determine if an orbit is ordered or chaotic, independent of the phase space.

Over the last years, several methods have been introduced, in order to determine and characterize the nature of orbits. One of the classical ways to do to that is by computing the maximal Lyapunov Characteristic Exponent (L.C.E) (see Lichtenberg and Lieberman, 1992). The main drawback with the L.C.E, is that this method is very time consuming as in most cases requires time periods of order of $10^{5}$ time units or even more, in order to obtain reliable results. This fact has lead researchers in search of new faster methods, in order to be able to distinguish between order and chaos in galactic dynamics and celestial mechanics. During the last years a variety of methods have been presented, in order to distinguish order from chaos. Among them it is worth mentioning the Fast Lyapunov Indicator - FLI (Froeschl\'{e} and Lega, 2000), the Generalized Alignment Index - GALI (Skokos et al., 2007), the Smaller Alignment Index - SALI (Skokos, 2001) and the $P(f)$ method used by Karanis and Vozikis (2008). Moreover, another interesting and fast method for distinguishing between order and chaos in Hamiltonian systems, is the method of the dynamical spectra of stretching numbers. Dynamical spectra have been frequently used in Galactic Dynamics, as fast and reliable indicators regarding the nature of motion (see Voglis et al., 1999; Caranicolas and Papadopoulos, 2007; Zotos, 2011a,b). The reader can find more details and theoretical explanations about the shape and behavior of the dynamical spectra, in the pioneer works of Froeschl\'{e} et al. (1993), Voglis and Contopoulos (1994) and Contopoulos et al. (1997).

The aim of the present article is to introduce and use a new dynamical parameter called the \textit{f}-indicator, in order to distinguish order from chaos in a class of Hamiltonian systems with two (2D) or three (3D) degrees of freedom, describing global or local motion in galaxies. There are two basic reasons for introducing the $f$-indicator: The first reason is that it is simple and easy to be applied in different kinds of dynamical systems. The second reason is that it provides fast and reliable results.

The article is organized as follows: In Section 2 we present the galactic potentials and give the definition of the $f$-indicator. In Section 3 we present results for the 2D systems. In the same Section we present results obtained using the L.C.E and the $S(c)$ dynamical spectrums in order to check the accuracy of the results given by the $f$-indicator. In Section 4 we study the nature of orbits in the 3D systems. A comparison with other dynamical parameters is also made in the same Section. We close with Section 5, where a discussion and the conclusions of this research are presented.

\section{The galactic potentials and the \textit{f}-indicator}

Two galactic potentials will be used in this research. The first is the logarithmic potential
\begin{equation}
V_G(x,y,z)=\frac{\upsilon_0^2}{2}ln\left[x^2-\lambda x^3 + \alpha y^2 + bz^2 + c_b^2\right].
\end{equation}
Potential (2.1) describes global motion in a triaxial elliptical galaxy. Here $\alpha, b$ are the flattening parameters, $c_b$ is the scale length of the bulge component, while the parameter $\lambda <<1$ introduces a small asymmetry in the system (see Binney and Tremaine, 1987). The parameter $\upsilon_0$ stands for the consistency of the galactic units. The choice of potential (1) is justified by the fact that triaxilities are common in elliptical galaxies (see e.g. Davies and Birkinshaw, 1988; Van Gorkom and Schiminovisch, 1997; Bak and Statler, 2000; Hibbard et al., 2001; Young, 2005).

The second potential is given by the equation
\begin{equation}
V_L(x,y,z)=\frac{\omega ^2}{2}\left(x^2+y^2+z^2\right)
-\epsilon\left[\beta\left(x^4+y^4+z^4\right) + 2\gamma\left(x^2y^2+y^2z^2+x^2z^2\right)\right].
\end{equation}
Here $\omega$ is the common unperturbed frequency of the oscillations along the $x, y$ and $z$ axis, $\beta, \gamma$ are parameters, while $\epsilon$ is the perturbation strength. Potential (2.2) is a 3D perturbed harmonic oscillator and describes local motion in the central parts of a galaxy. Such local 2D or 3D potentials, are products of expansion of global galactic potentials in a Taylor series near a stable equilibrium point and have been extensively used in order to describe local motion in galaxies (see Deprit and Elipe, 1991; Caranicolas, 1994; Elipe and Deprit, 1999; Elipe, 2000; Arribas et al., 2006; Zotos, 2010).

The outcomes of this research are based on the numerical integration of the equations of motion
\begin{eqnarray}
\ddot{x}&=&-\frac{\partial \ V(x,y,z)}{\partial x}, \nonumber \\
\ddot{y}&=&-\frac{\partial \ V(x,y,z)}{\partial y}, \nonumber \\
\ddot{z}&=&-\frac{\partial \ V(x,y,z)}{\partial z},
\end{eqnarray}
where $V(x,y,z)$ is one of the $V_G(x,y,z)$ or $V_L(x,y,z)$ and the dot indicates derivatives with respect to time. The corresponding Hamiltonian reads
\begin{equation}
H=\frac{1}{2}\left(p_x^2+p_y^2+p_z^2\right)+V(x,y,z)=C \ \ \ ,
\end{equation}
where $p_x, p_y, p_z$ are the momenta per unit mass conjugate to $x, y$ and $z$ respectively, while $C$ is the numerical value of the Hamiltonian. We use $C=E$ or $C=h$, for the global or the local potential respectively.

In this paper, we use a system of galactic units, where the unit of length is 1kpc, the unit of mass is 2.325 $\times$ $10^7$ M$_{\odot}$ and the unit of time is 0.97748 $\times$ $10^8$ yr. The velocity and the angular velocity units are 10 km/s and 10 km/s/kpc respectively, while $G$ is equal to unity. The energy unit (per unit mass) is 100 (km/s)$^2$. In the above units we use the values: $\upsilon_0=15, c_b =2.5, \alpha=1.5, b=1.7, \omega =0.6, \beta =0.2, \epsilon =0.4$, while $\lambda$ and $\gamma$ are treated as parameters.

All numerical calculations of the present research, are based on the numerical integration of the equations of motion (2.3), which was made using a Bulirsh-St\"{o}er integration routine in Fortran 95, with double precision in all subroutines. The accuracy of our calculations was checked by the constancy of the energy integral (2.4), which was conserved up to the twelfth significant figure. We have constructed very accurate integration routines, in order to minimize as much as possible the errors (noise). For all numerical calculations of the present research, the errors were of order of $10^{-14}$ or even smaller and therefore are negligible.

The $f$-indicator for a 3D potential is defined as
\begin{equation}
\textit{f}_3=\frac{E_z}{E_x+E_y} \ \ \ ,
\end{equation}
where
\begin{eqnarray}
E_x&=&\frac{1}{2}p_x^2+V(x,y=0,z=0), \nonumber \\
E_y&=&\frac{1}{2}p_y^2+V(x=0,y,z=0), \nonumber \\
E_z&=&\frac{1}{2}p_z^2+V(x=0,y=0,z),
\end{eqnarray}
are the energies of the separable system $(\alpha _1=0)$, along the $x, y$ and $z$ axis. For the local potential $V_L$ we have
\begin{eqnarray}
E_x&=&\frac{1}{2}\left(p_x^2+\omega ^2 x^2\right)-\epsilon \beta x^4, \nonumber \\
E_y&=&\frac{1}{2}\left(p_y^2+\omega ^2 y^2\right)-\epsilon \beta y^4, \nonumber \\
E_z&=&\frac{1}{2}\left(p_z^2+\omega ^2 z^2\right)-\epsilon \beta z^4,
\end{eqnarray}
while for the global potential $V_G$ we take
\begin{eqnarray}
E_x&=&  \frac{1}{2}p_x^2+V(x,y=0,z=0)=\frac{1}{2}p_x^2+\frac{\upsilon _0^2}{2}ln\left(x^2-\lambda x^3 + c_b^2\right), \nonumber \\
E_y&=&\frac{1}{2}p_y^2+V(x=0,y,z=0)=\frac{1}{2}p_y^2+\frac{\upsilon _0^2}{2}ln\left(\alpha y^2 + c_b^2\right), \nonumber \\
E_z&=&\frac{1}{2}p_z^2+V(x=0,y=0,z)=\frac{1}{2}p_z^2+\frac{\upsilon _0^2}{2}ln\left(bz^2 + c_b^2\right).
\end{eqnarray}

For a 2D potential the $f$-indicator is defined as
\begin{equation}
\textit{f}_2=\frac{E_x}{E_y} \ \ \ .
\end{equation}

In order to avoid confusion we would like to draw the attention of the reader to the following point: the $f$-indicator is defined by equations (2.5) and (2.9), as if the potentials used were separable. But in all cases, $Ex$, $Ey$ and $Ez$ and consequently the $f$-indicator, are computed not for the separable potentials but for the potentials given by relations (2.1) and (2.2), where the coupling parameters $\alpha, b$ and $\gamma$ are always present and different from zero. This means that $Ex$, $Ey$ and $Ez$ entering equations (2.5) or (2.9), are not any more integrals of motion in all presented calculations. On this basis, there is a coupling of the energies of the test particle along the $x,y$ and $z$ direction. This coupling, although hidden, is entering in the formulae given by equations (2.5) and (2.9). It is this coupling that defines the regular or chaotic character of orbit.

A second interesting point is that the above method can be applied to a wide range of potentials, such as potential made up of harmonics oscillators or global mass models and logarithmic galactic potentials (see Binney and Tremaine, 1987, pp 42). The author would like to make clear that there might be cases of more complex potentials such as those describing spiral arms or binary systems, where the method of the $f$-indicator is not easy to be applied.

Before closing this Section, we would like to recall the definition of the $S(c)$ spectrum. The parameter $c$ is defined as
\begin{equation}
c_i=\frac{x_i-p_{xi}}{p_{yi}} \ \ \ ,
\end{equation}
where $\left(x_i,p_{xi},p_{yi}\right)$ are the successive values $\left(x_i,p_{xi},p_{yi}\right)$ on the Poincar\'{e} phase plane. The dynamical spectrum of the parameter $c$ is its distribution function
\begin{equation}
S(c)=\frac{\Delta N(c)}{N \Delta c} \ \ \ ,
\end{equation}
where $\Delta N(c)$ are the number of the parameters $c$ in the interval $\left(c, c+ \Delta c\right)$, after $N$ iterations. The $S(c)$ spectrum can be used both in 2D and 3D potentials. Note that, the coupling of the third component $z$, is hidden (see also Caranicolas and Zotos, 2010), but in any case it affects the values of $x,p_x$ and $p_y$ entering equation (2.10).

\section{Results for the 2D systems}

Let us now proceed to study the character of motion in the corresponding 2D potentials. First we shall present results for the global 2D potential
\begin{equation}
V_G(x,y)=\frac{\upsilon _0^2}{2}ln\left[x^2 - \lambda x^3 + \alpha y^2 + c_b^2\right] \ \ \ .
\end{equation}

The corresponding Hamiltonian to potential (3.1) writes
\begin{equation}
H_{G2}=\frac{1}{2}\left(p_x^2+p_y^2\right)+V_G(x,y)=E_2 \ \ \ ,
\end{equation}
where $E_2$ is the numerical value of the Hamiltonian.

A simple qualitative way of revealing the dynamical structure of a Hamiltonian system, is by plotting the successive intersections of the orbits with a Poincar\'{e} surface of section (P.S.S). This method, has been extensively applied to Hamiltonian systems with two degrees of freedom, as in these systems the P.S.S is a two dimensional plane. In dynamical systems with three degrees of freedom, however, the P.S.S is four dimensional and the behavior of the orbits cannot be easily visualized. One way to overcome this drawback, is to project the P.S.S to space with lower dimensions (Caranicolas and Zotos, 2009). However, even these projections are often very complicated and difficult to interpret.

Figure 1a shows the $(x,p_x), (y=0,p_y>0)$ phase plane, when $\lambda = 0.01$ and $E_2 = 487$. As one can observe, a large part of the phase plane is covered by chaotic orbits. The regular regions are confined around the two stable periodic points on the $x$ axis. These points represent two identical stable periodic orbits around the origin traversed in opposite directions, characteristics of the 1:1 resonance. There are also regular regions in the outer parts of the phase plane. The corresponding orbits are quasi periodic orbits characteristics of the 2:3 resonance and  box orbits. There are also some small islands produced by secondary resonances. These secondary resonances represent complicated orbits of the dynamical system (see Fig. 3a). Figure 1b is similar to Fig. 1a but when $\lambda = 0.02$ and $E_2 = 494$. Here the chaotic region is larger. The regular regions around the periodic points on the $x$ axis have decreased, while there are not any box orbits. Secondary resonances are also present.

\begin{figure}[htpb]
  \centering
  \subfloat[$\lambda=0.01,E_2=487$]{\includegraphics[angle=270, scale=0.3]{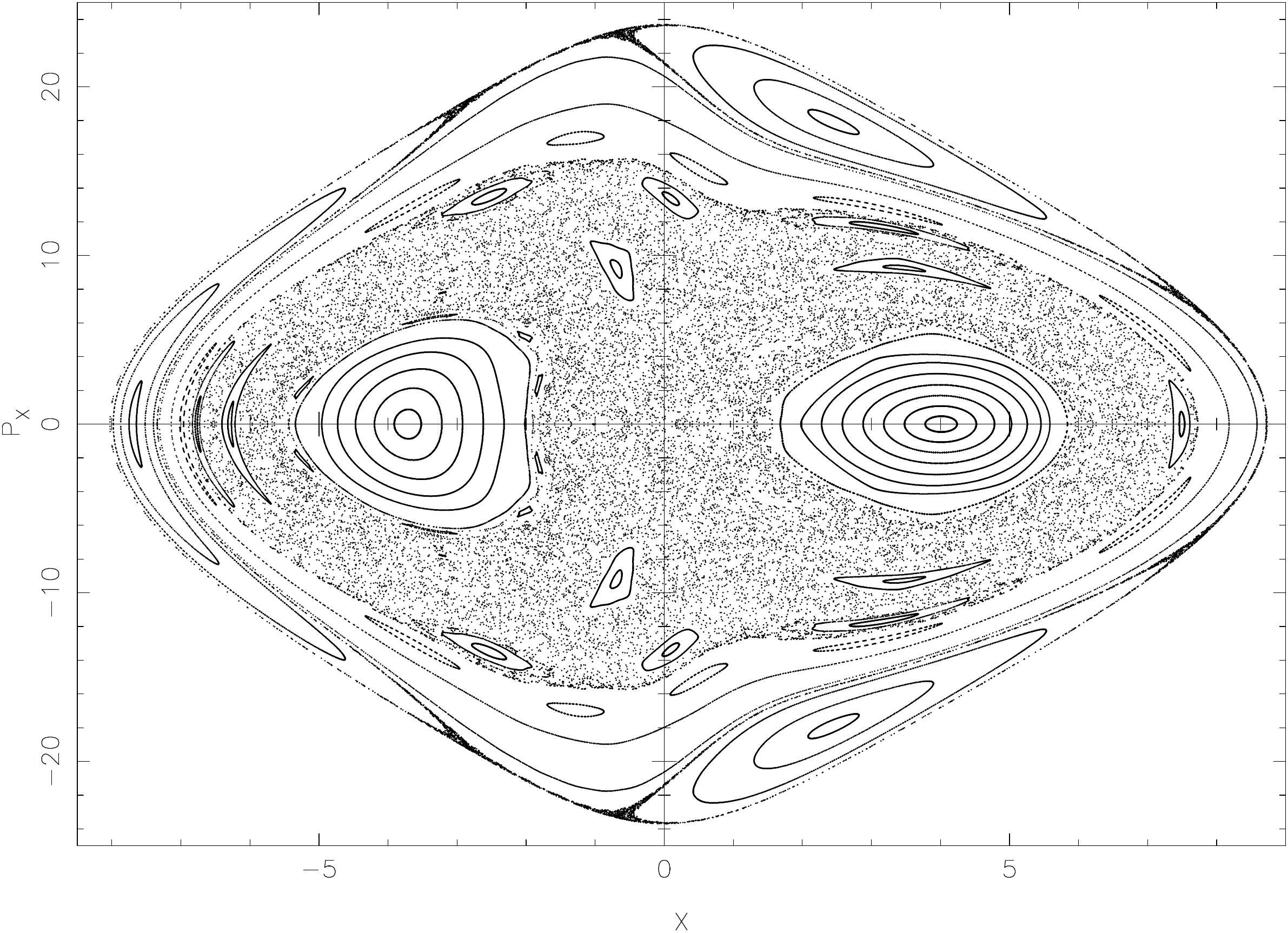}\label{Fig-1a}}
\hspace*{1cm}
  \subfloat[$\lambda =0.02, E_2=494$.]{\includegraphics[angle=270, scale=0.3]{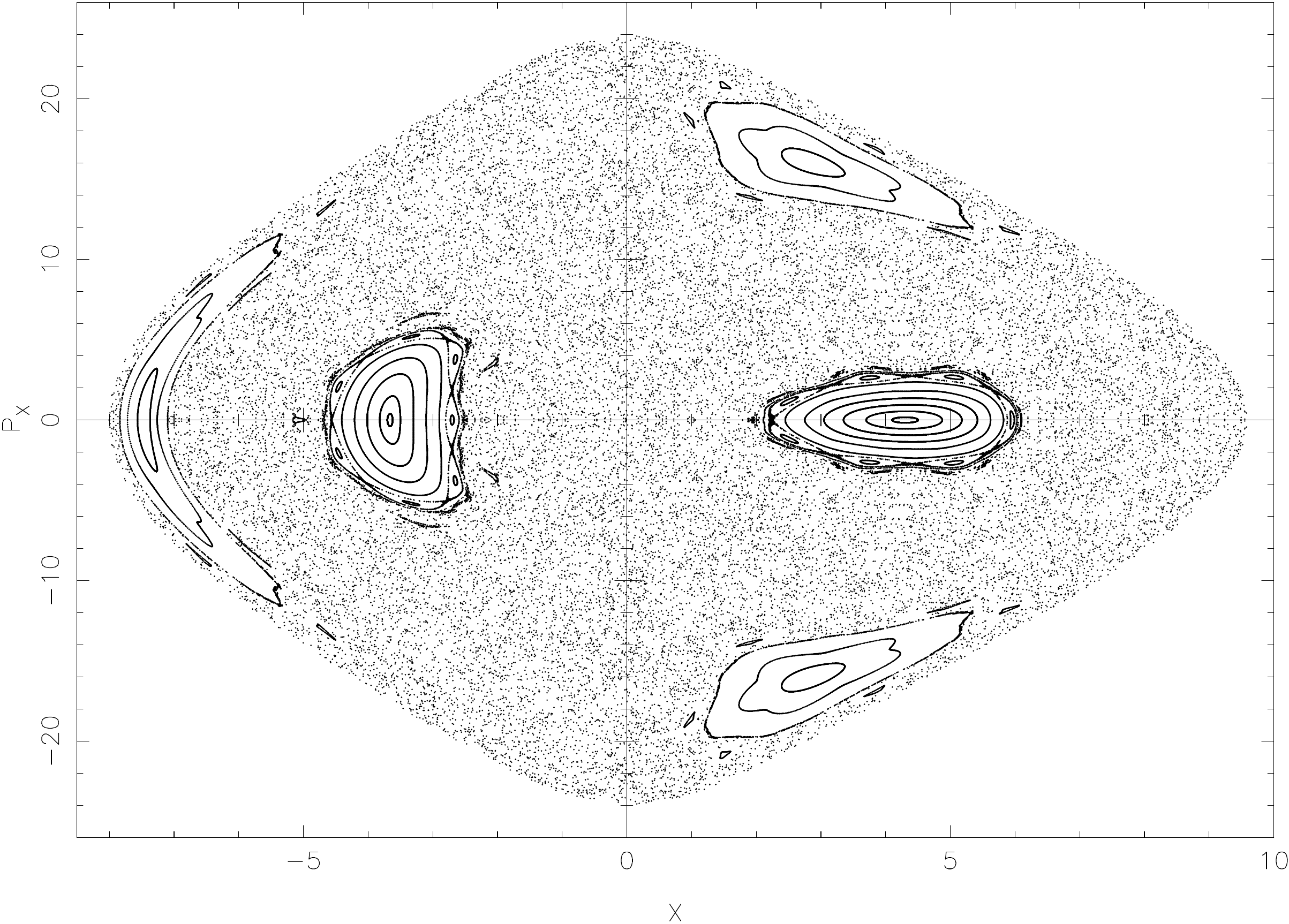}\label{Fig-1b}}
\caption{The $(x,p_x)$ phase plane for the global potential.}
\label{fig:two-subfigures-one}
\end{figure}

The local 2D potential is
\begin{equation}
V_L(x,y)=\frac{\omega ^2}{2}\left(x^2+y^2\right)
-\epsilon \left[\beta \left(x^4+y^4\right) + 2\gamma x^2y^2\right] \ \ \ ,
\end{equation}
while the corresponding  Hamiltonian is
\begin{equation}
H_{L2}=\frac{1}{2}\left(p_x^2+p_y^2\right)+V_L(x,y)=h_2 \ \ \ ,
\end{equation}
where $h_2$ is the numerical value of the Hamiltonian. The results obtained from the study of the 2D local system, will be used in order to help us to understand and reveal the structure of the more complicated phase space of the 3D local system, which will be presented in the following section.

The corresponding phase planes are shown in Figures 2a-b. The value of $h_2$ is 0.10125 in both figures, while $\gamma = -0.2$ for Fig. 2a and $\gamma = -0.8$ for Fig. 2b. In Fig. 2a the phase plane is covered by regular orbits except for a small chaotic layer near the separatrix, while in Fig. 2b the larger part of the phase plane is covered by chaotic orbits. Regular regions are observed near the stable periodic points on the $x$ and $p_x$ axis. The corresponding orbits are quasi periodic characteristics of the 1:1 resonance.

\begin{figure}[htbp]
\begin{center}
  \subfloat[ $\gamma=-0.2$.]{\includegraphics[angle=270, scale=0.3]{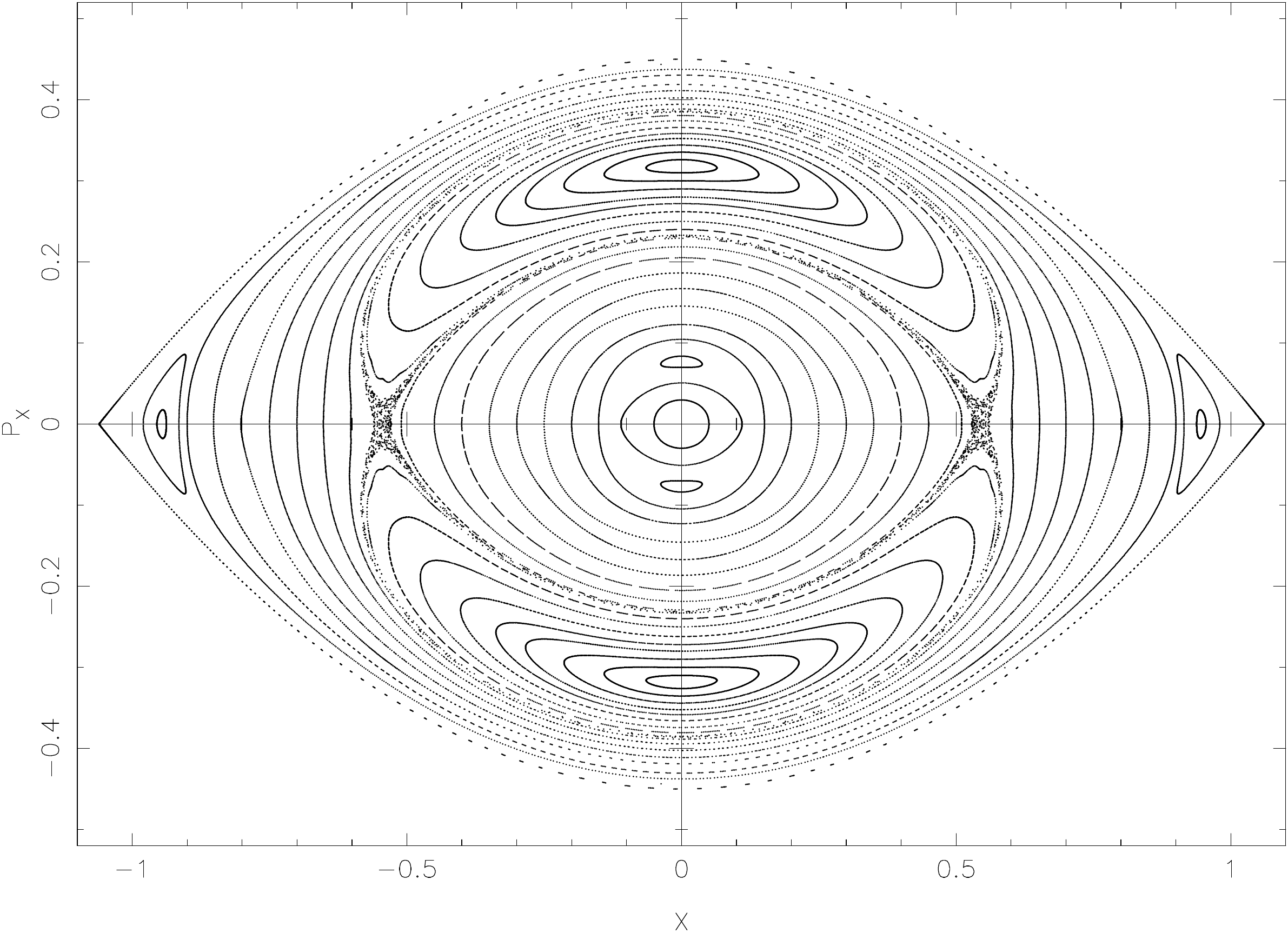}\label{Fig-2a}}
\hspace*{1cm}
  \subfloat[$\gamma=-0.8$.]{\includegraphics[angle=270, scale=0.3]{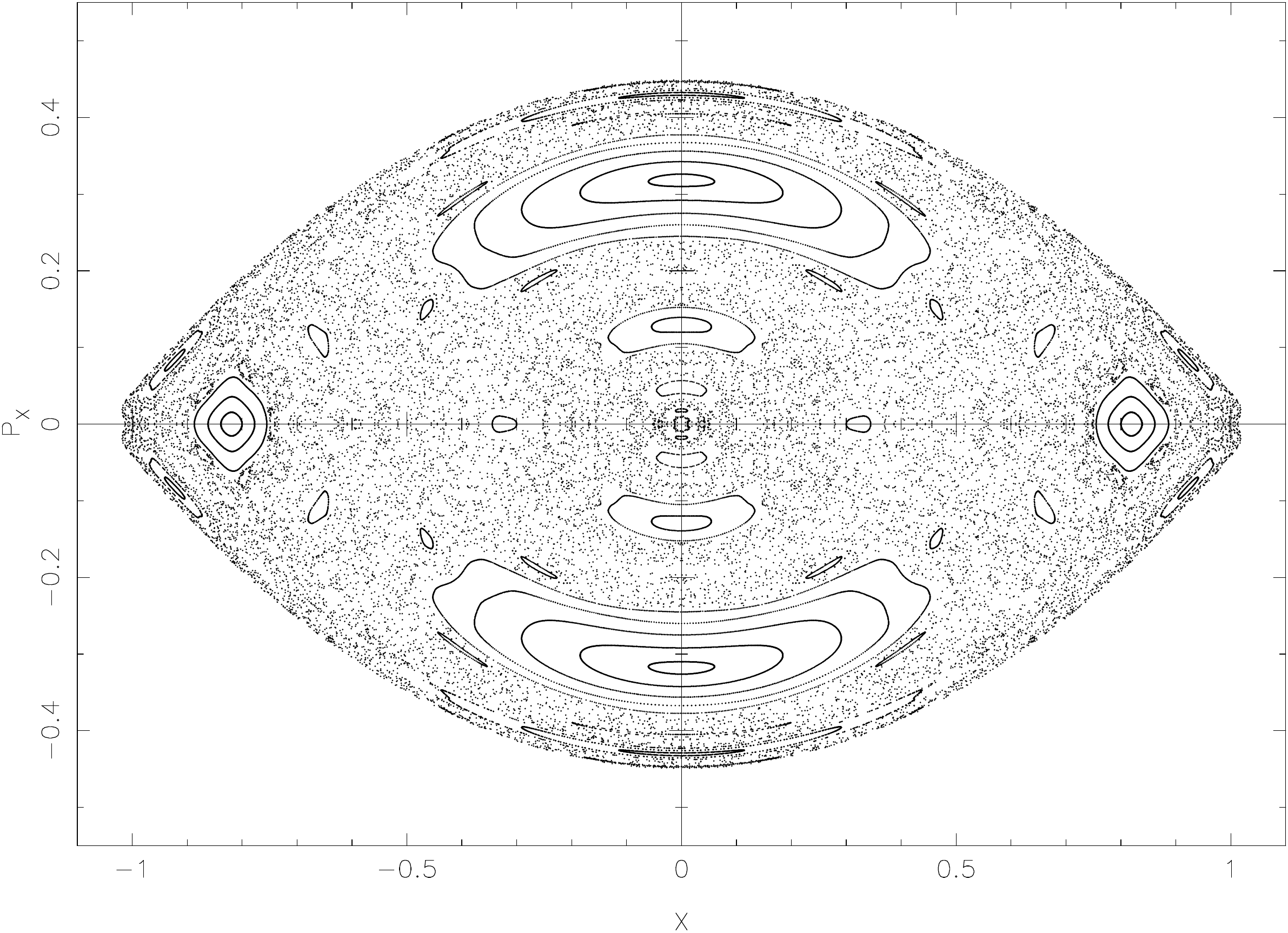}\label{Fig-2b}}
\end{center}
\caption{The $(x,p_x)$ phase plane for the local potential when $h_2=0.10125$.}
\label{fig:two-subfigures-one}
\end{figure}

In the following we shall study the character of orbits in the 2D potential using three different criteria. The first is the L.C.E (see Lichtenberg \& Lieberman, 1992), the second is the $S(c)$ spectrum (see Caranicolas and Papadopoulos, 2007; Caranicolas and Zotos, 2010) and the third is the new $f$-indicator. Figure 3a-d shows results for an orbit in the 2D global potential (3.1). This orbit is a characteristic example of the 3:4 resonance. The initial conditions are: $x_0=7.47, p_{x0}=0$, while for all orbits $y_0=0$ and the value of $p_{y0}$ is always found using the Hamiltonian (3.2). The values of all other parameters are as in Fig. 1a. In Fig. 3a we observe a quasi periodic. One can see in Fig. 3b that the L.C.E vanishes as expected, the $S(c)$ dynamical spectrum produces four $U$ type spectra showing that the orbit produces four islands on the phase plane. Looking at the plot of the evolution of the $f_2$-indicator with time shown in Fig. 3d, we observe a quasi periodic profile with almost symmetric peaks. This suggests, that the corresponding orbit is regular. The integration time of orbit shown in Fig. 3a, is 100 time units, while for the $S(c)$ spectrum is $2 \times 10^4$ time units.

\begin{figure}[htbp]
\centering
\subfloat[ A quasi periodic orbit in the 2D global potential.]{\includegraphics[angle=0, scale=0.45]{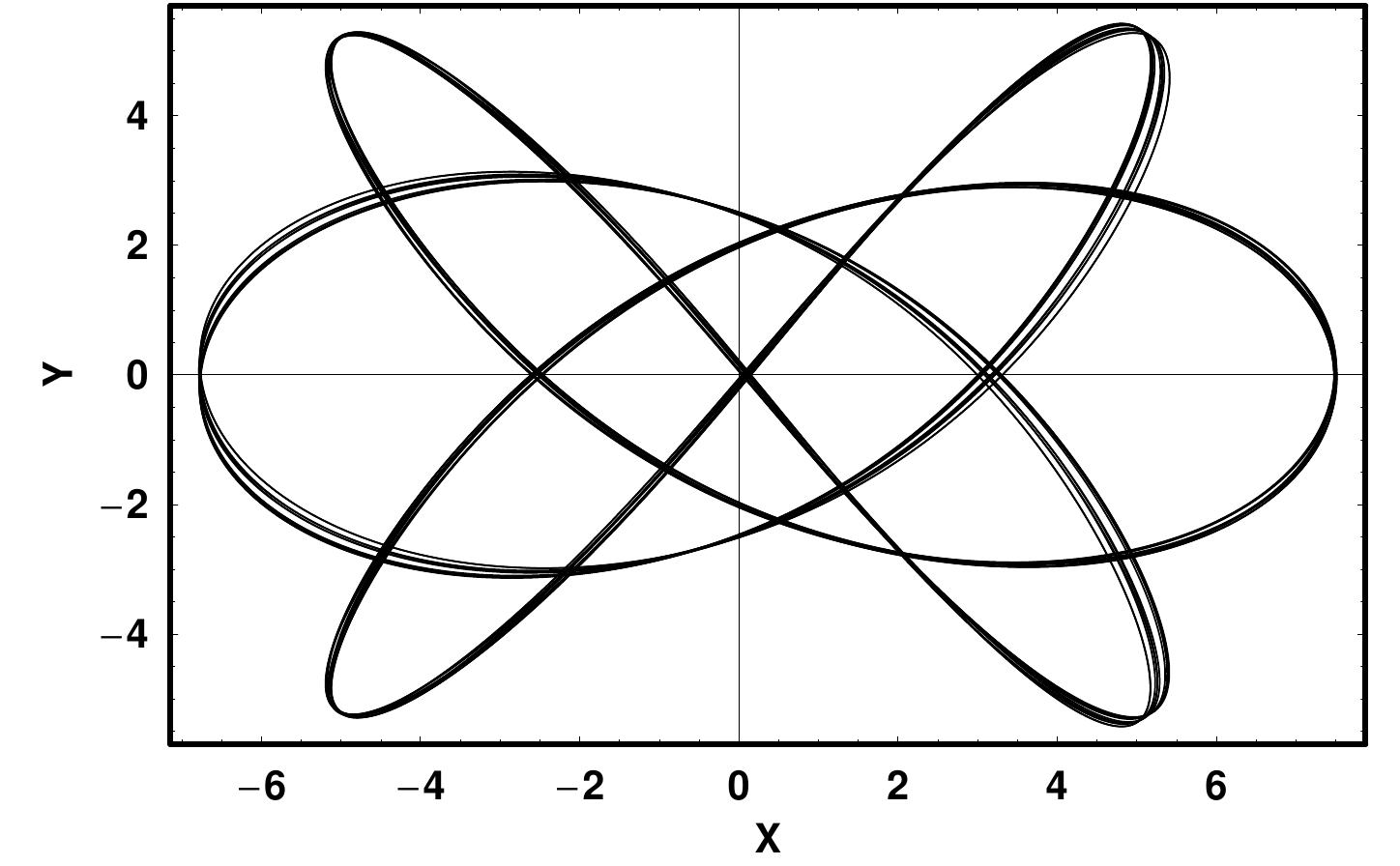}\label{Fig-3a}}
\hspace*{1cm}
\subfloat[ The corresponding L.C.E.]{\includegraphics[angle=0, scale=0.45]{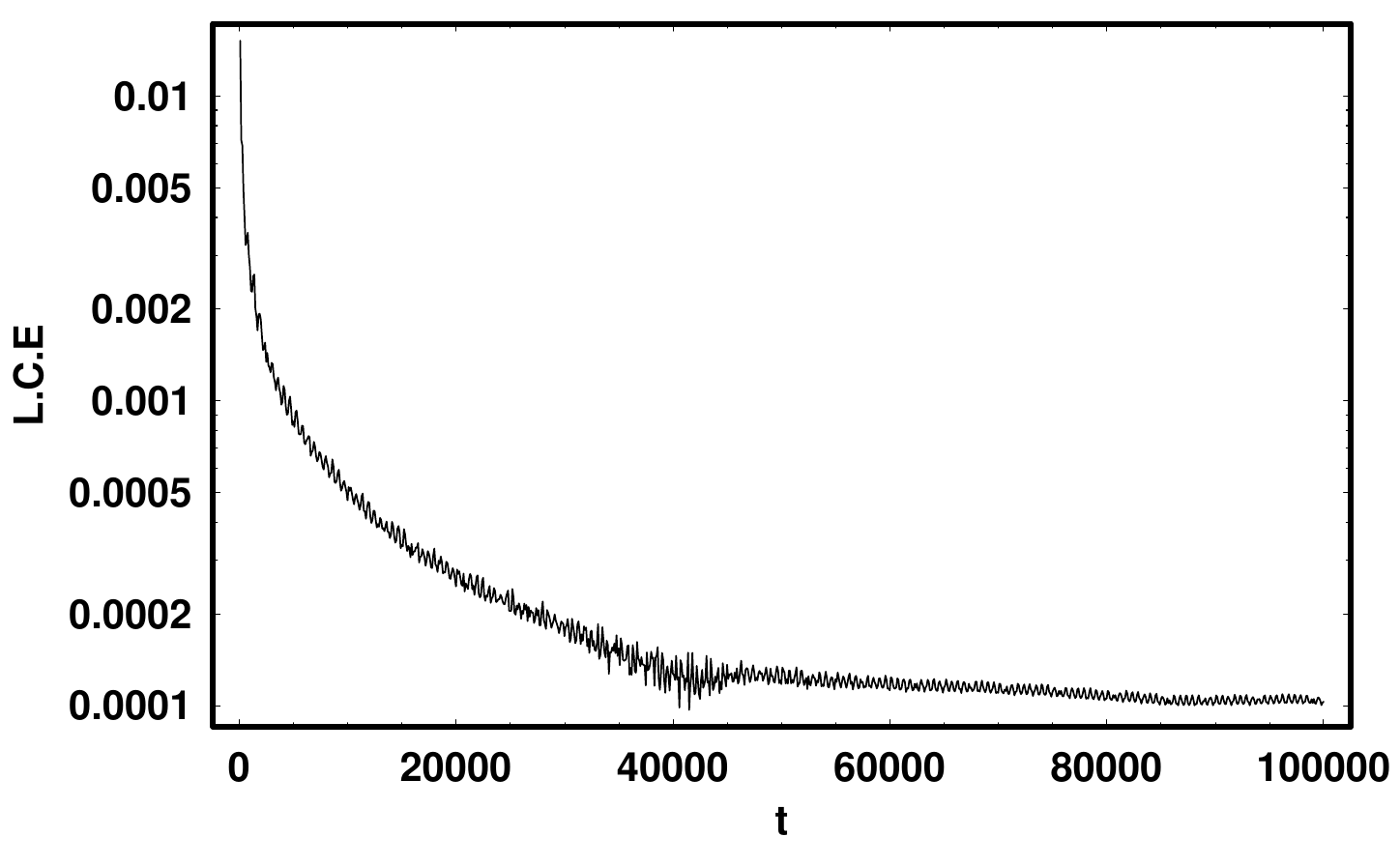}\label{Fig-3b}}\\
\subfloat[ The $S(c)$ spectrum]{\includegraphics[angle=0, scale=0.45]{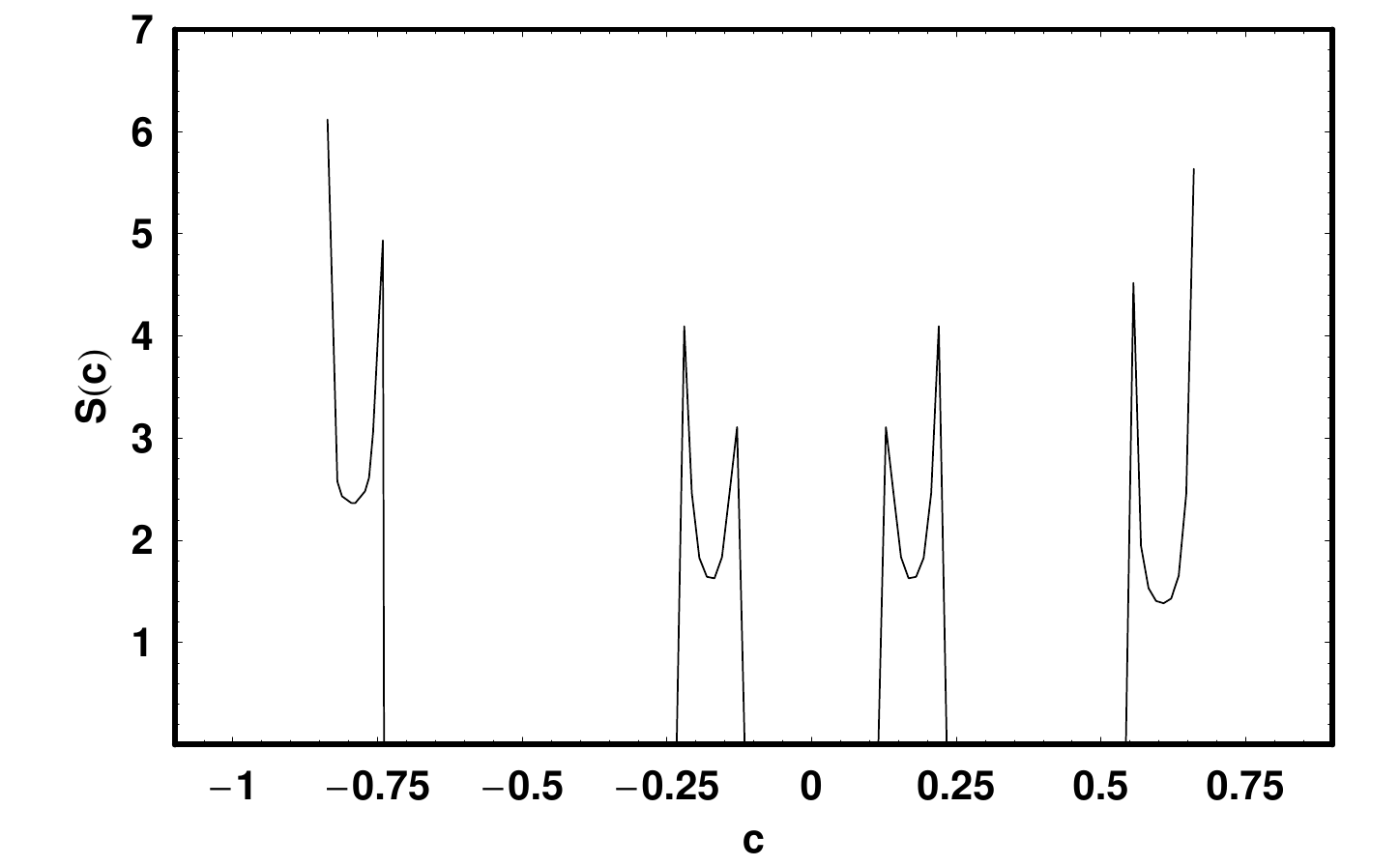}\label{Fig-3c}}
\hspace*{1cm}
\subfloat[ The corresponding $f_2$-indicator.]{\includegraphics[angle=0, scale=0.45]{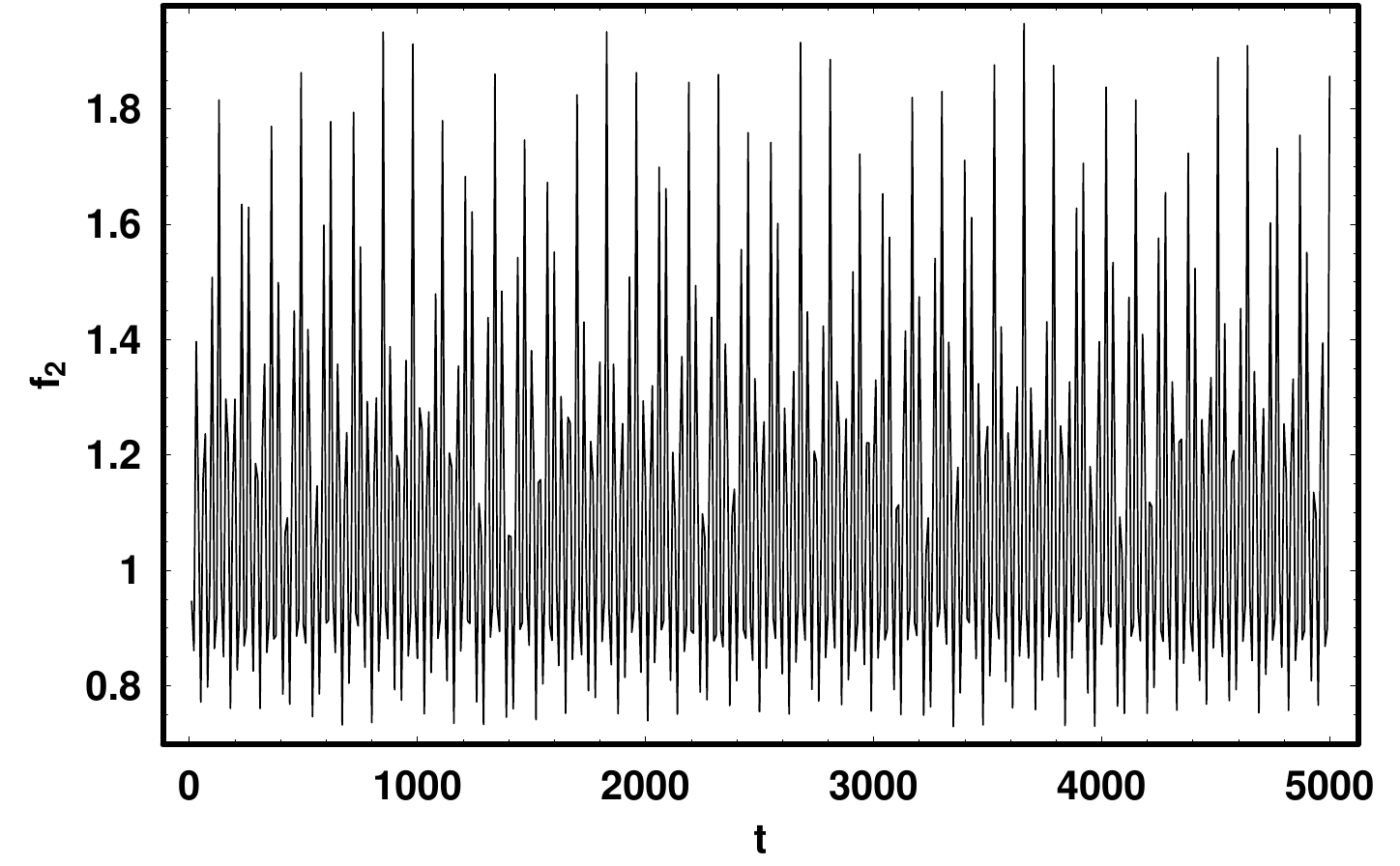}\label{Fig-3d}}
\caption{}
\end{figure}

\begin{figure}[htbp]
\centering
\subfloat[ A quasi periodic orbit in the 2D global potential.]{\includegraphics[angle=0, scale=0.45]{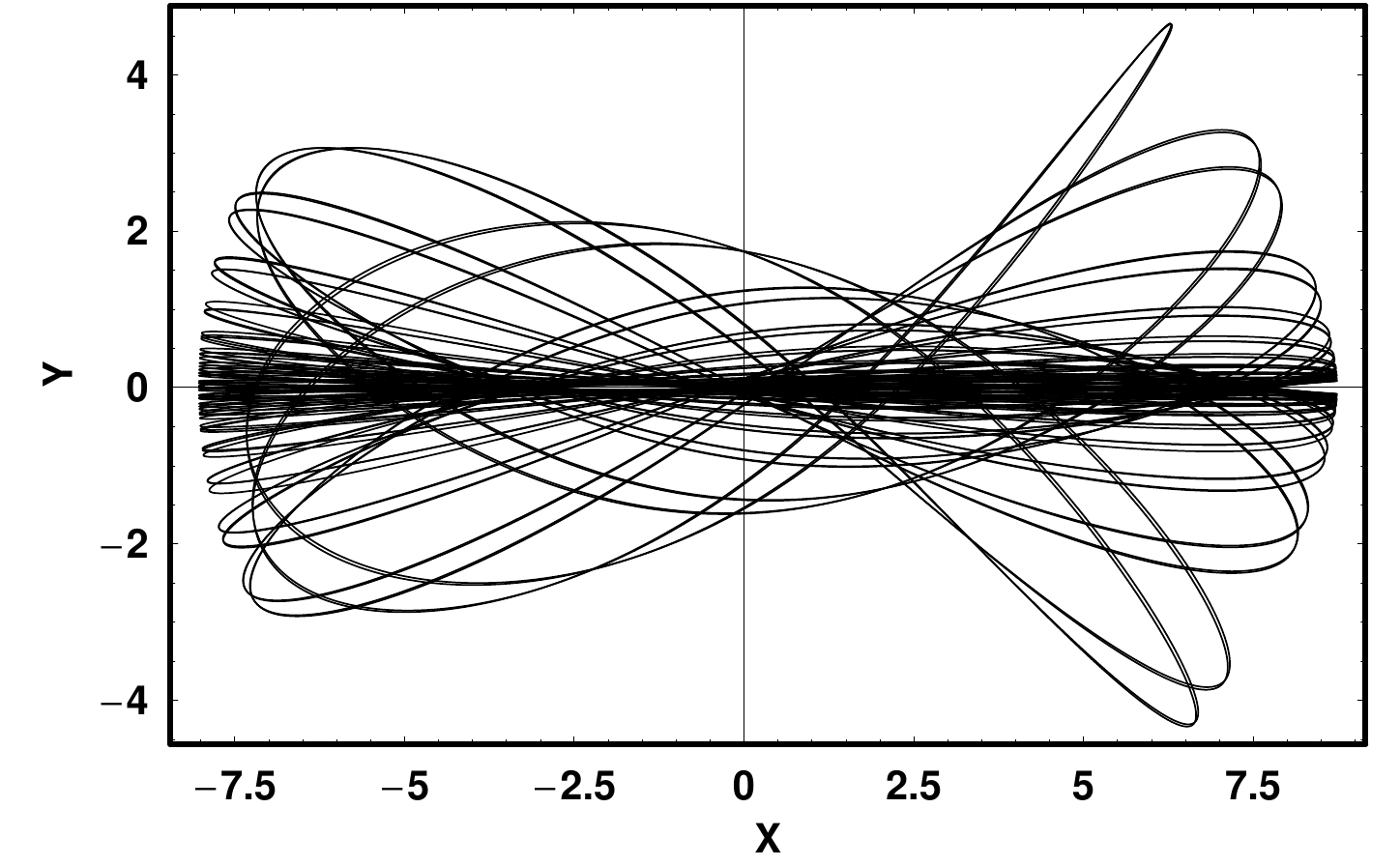}\label{Fig-4a}}
\hspace*{1cm}
\subfloat[ The corresponding L.C.E.]{\includegraphics[angle=0, scale=0.45]{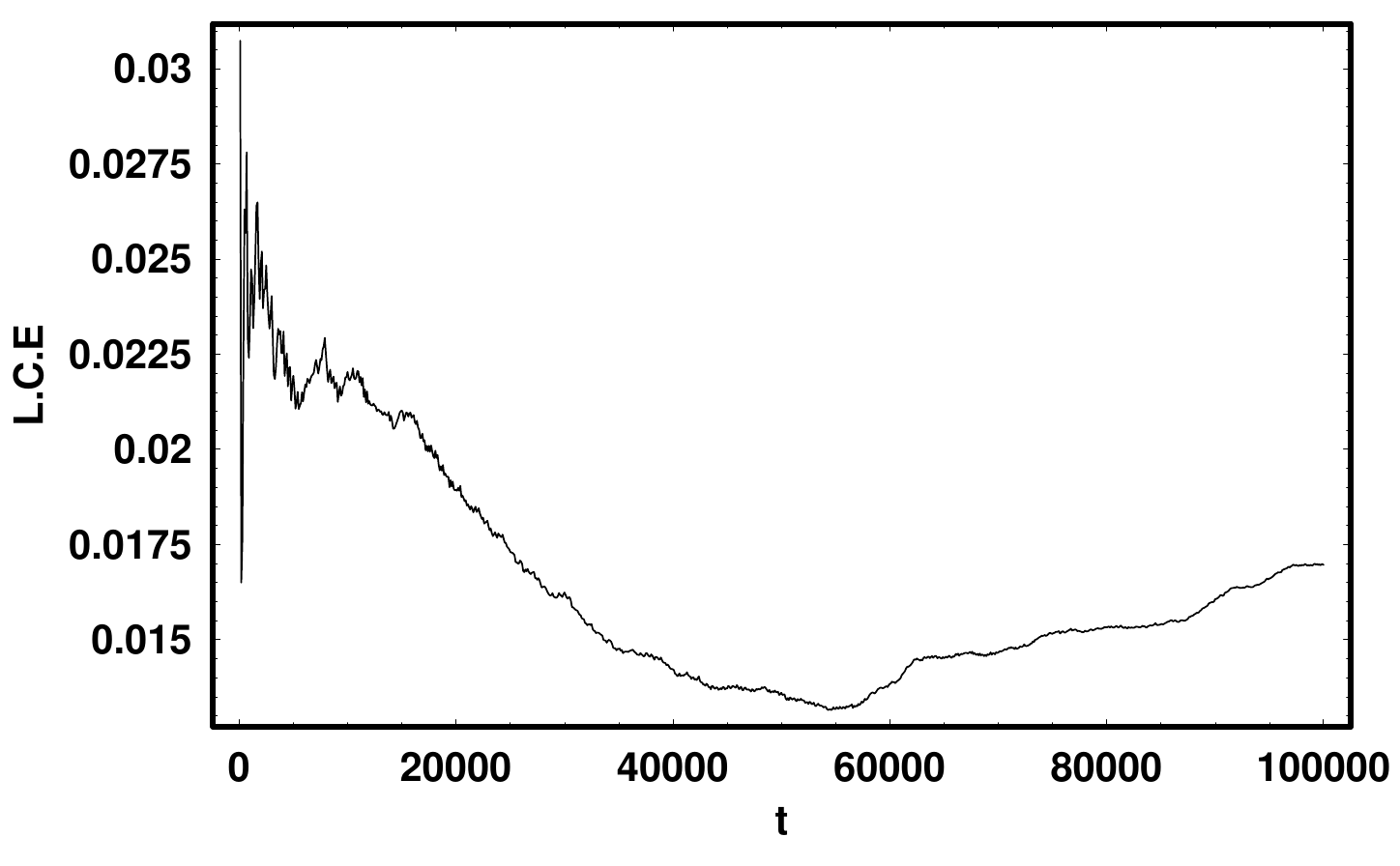}\label{Fig-4b}}\\
\subfloat[ The $S(c)$ spectrum]{\includegraphics[angle=0, scale=0.45]{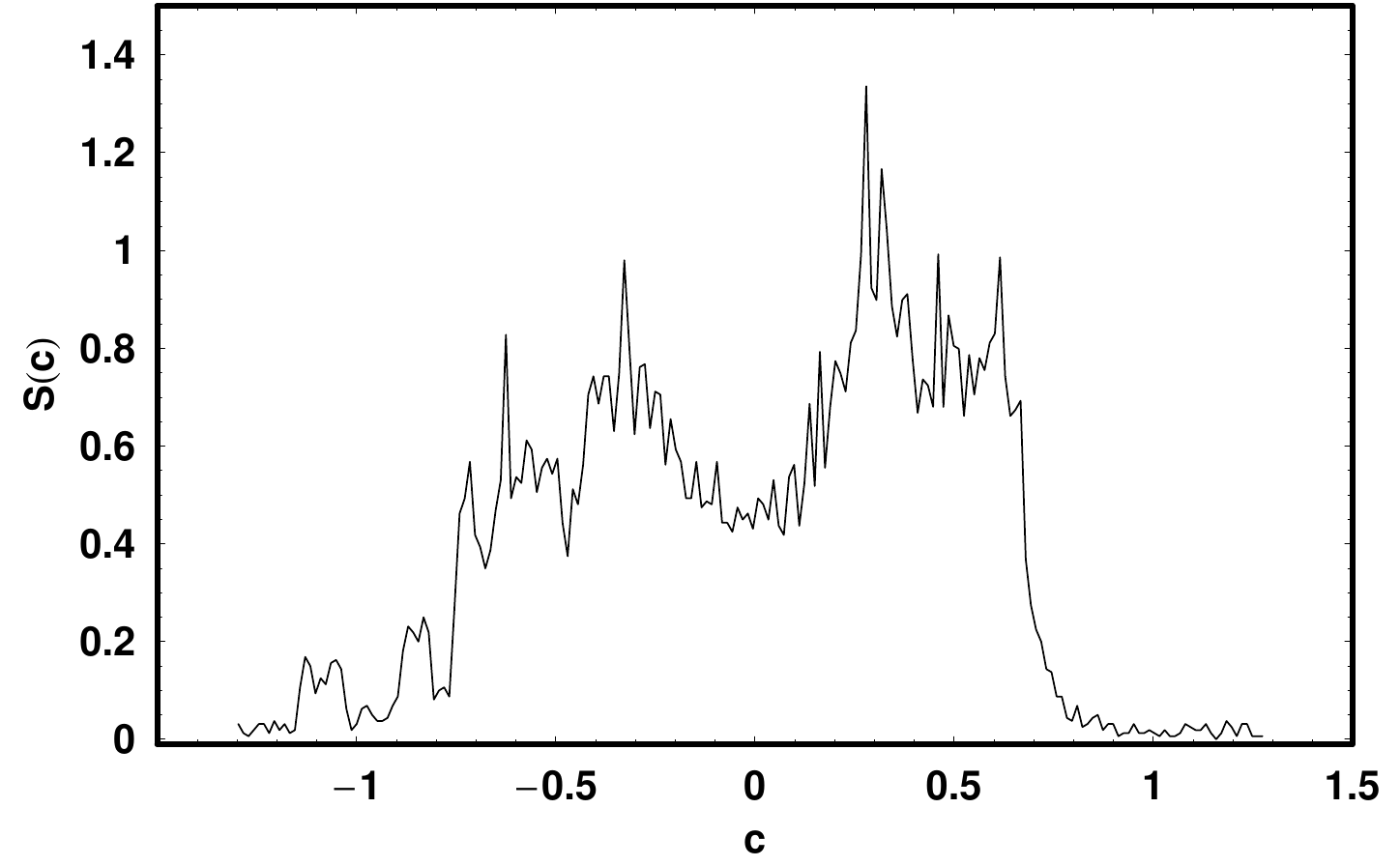}\label{Fig-4c}}
\hspace*{1cm}
\subfloat[ The corresponding $f_2$-indicator.]{\includegraphics[angle=0, scale=0.45]{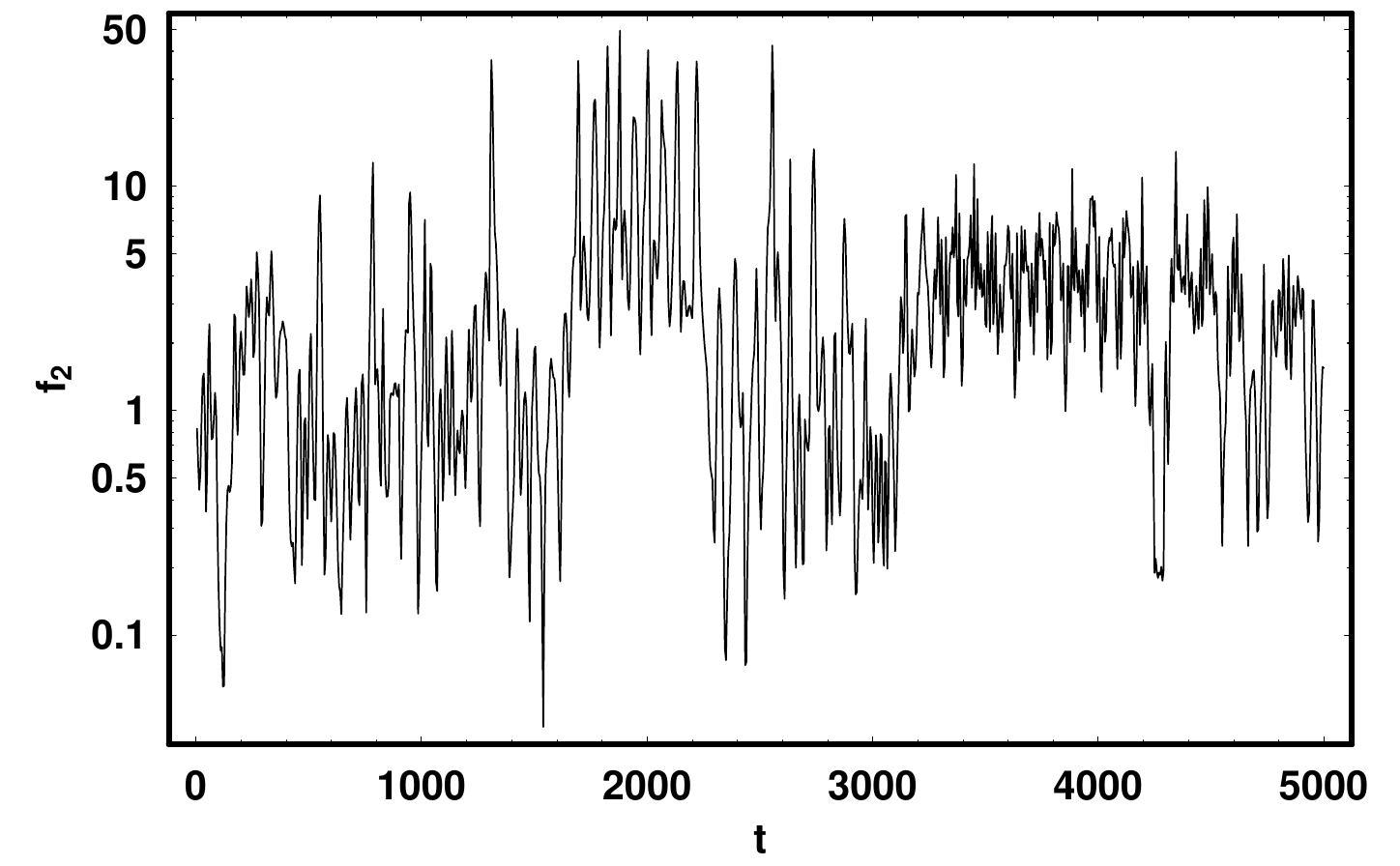}\label{Fig-4d}}
\caption{For a borderline chaotic orbit in the 2D global potential. See text for details.}
\end{figure}

\begin{figure}[htbp]
\centering
\subfloat[ A regular orbit in the 2D local potential.]{\includegraphics[angle=0, scale=0.45]{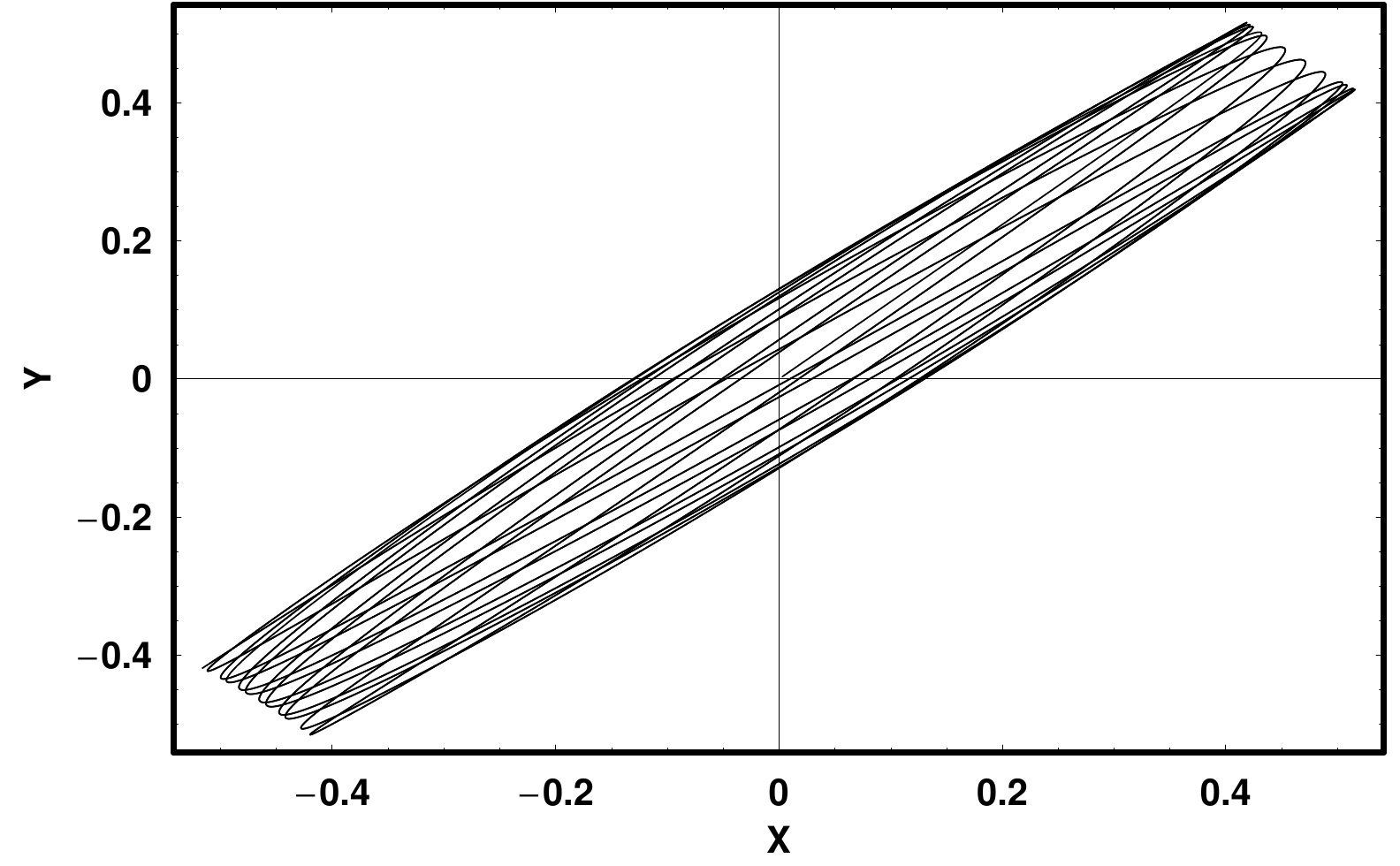}\label{Fig-5a}}
\hspace*{1cm}
\subfloat[ The corresponding L.C.E.]{\includegraphics[angle=0, scale=0.45]{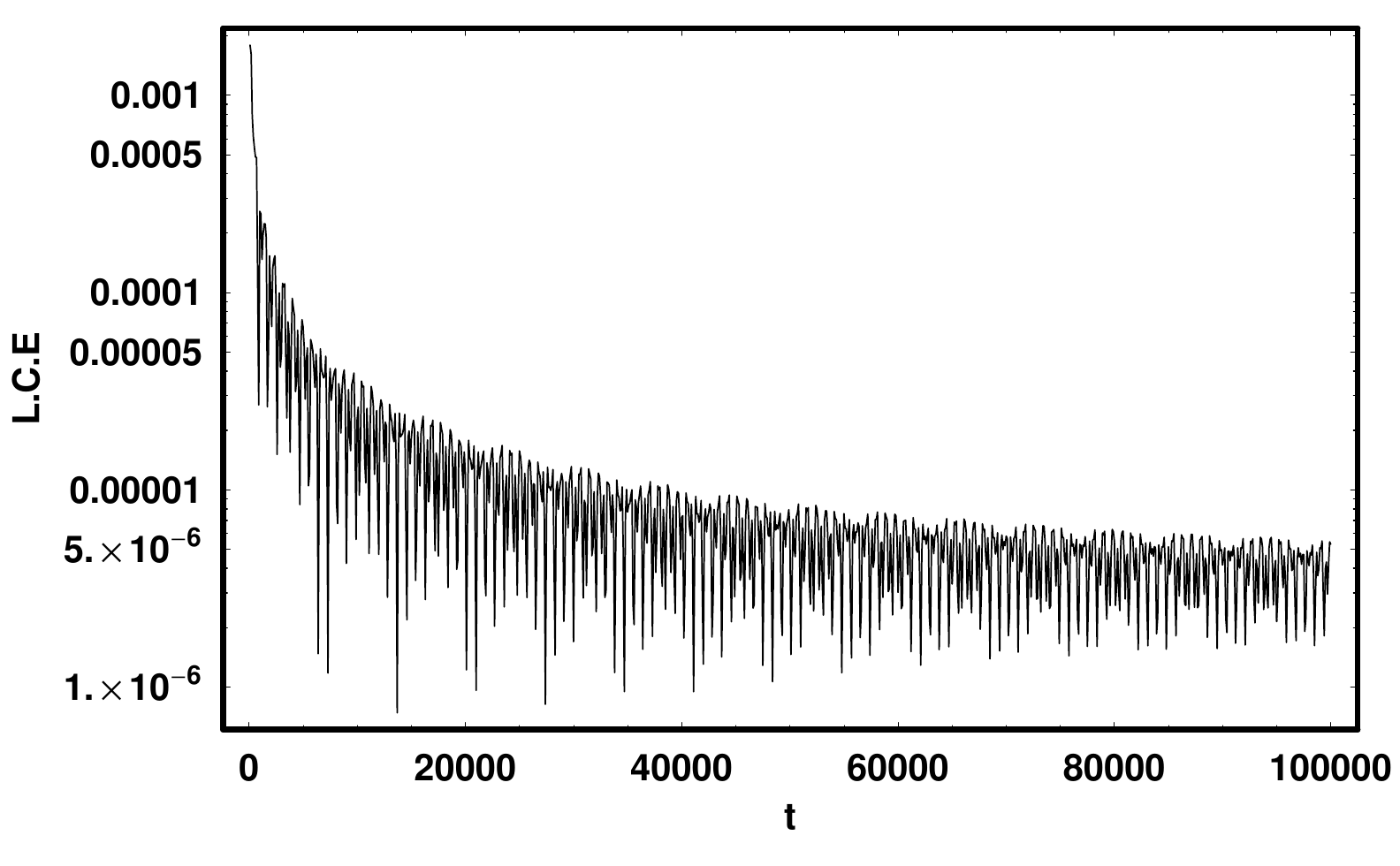}\label{Fig-5b}}\\
\subfloat[ The $S(c)$ spectrum]{\includegraphics[angle=0, scale=0.45]{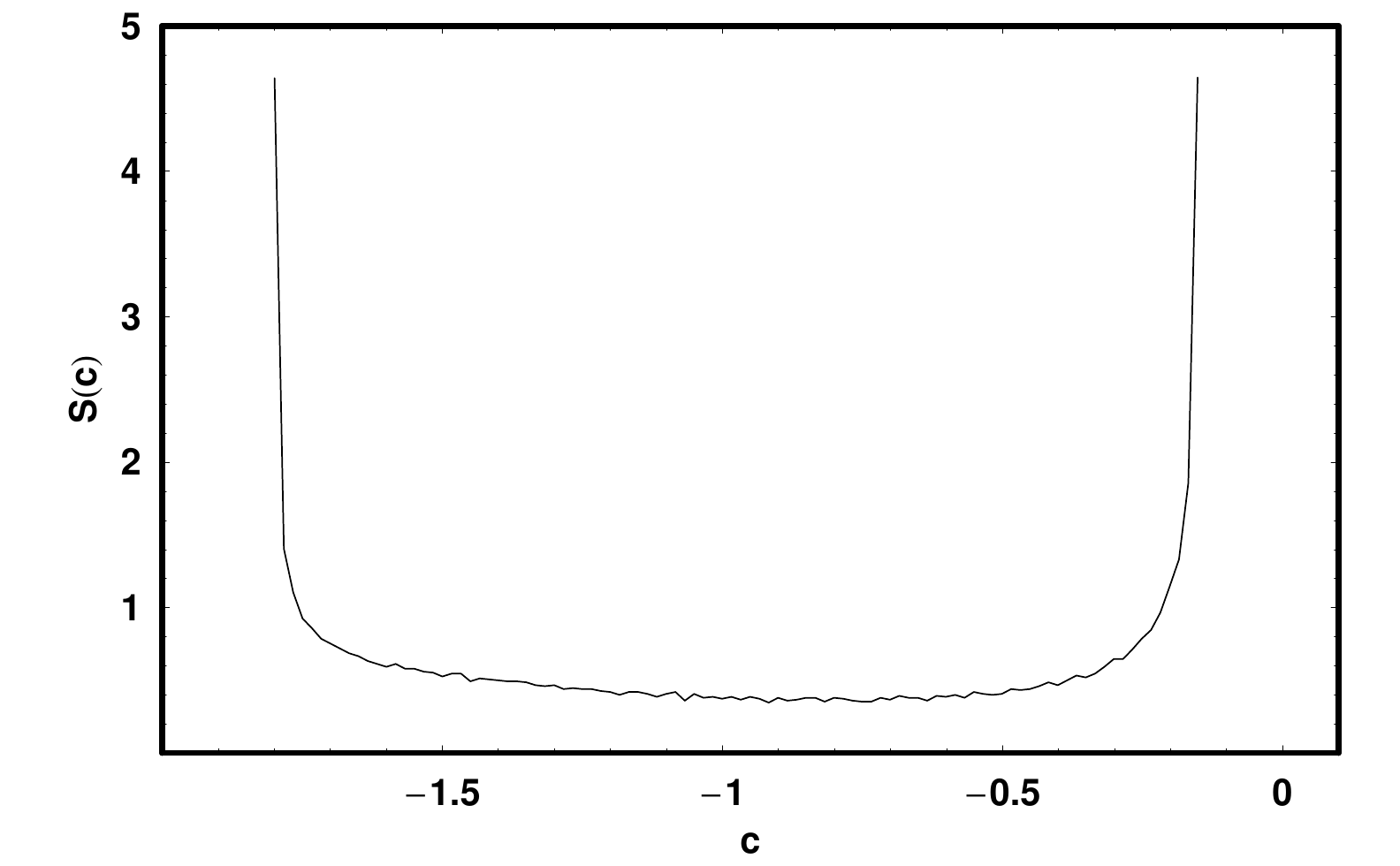}\label{Fig-5c}}
\hspace*{1cm}
\subfloat[ The corresponding $f_2$-indicator.]{\includegraphics[angle=0, scale=0.45]{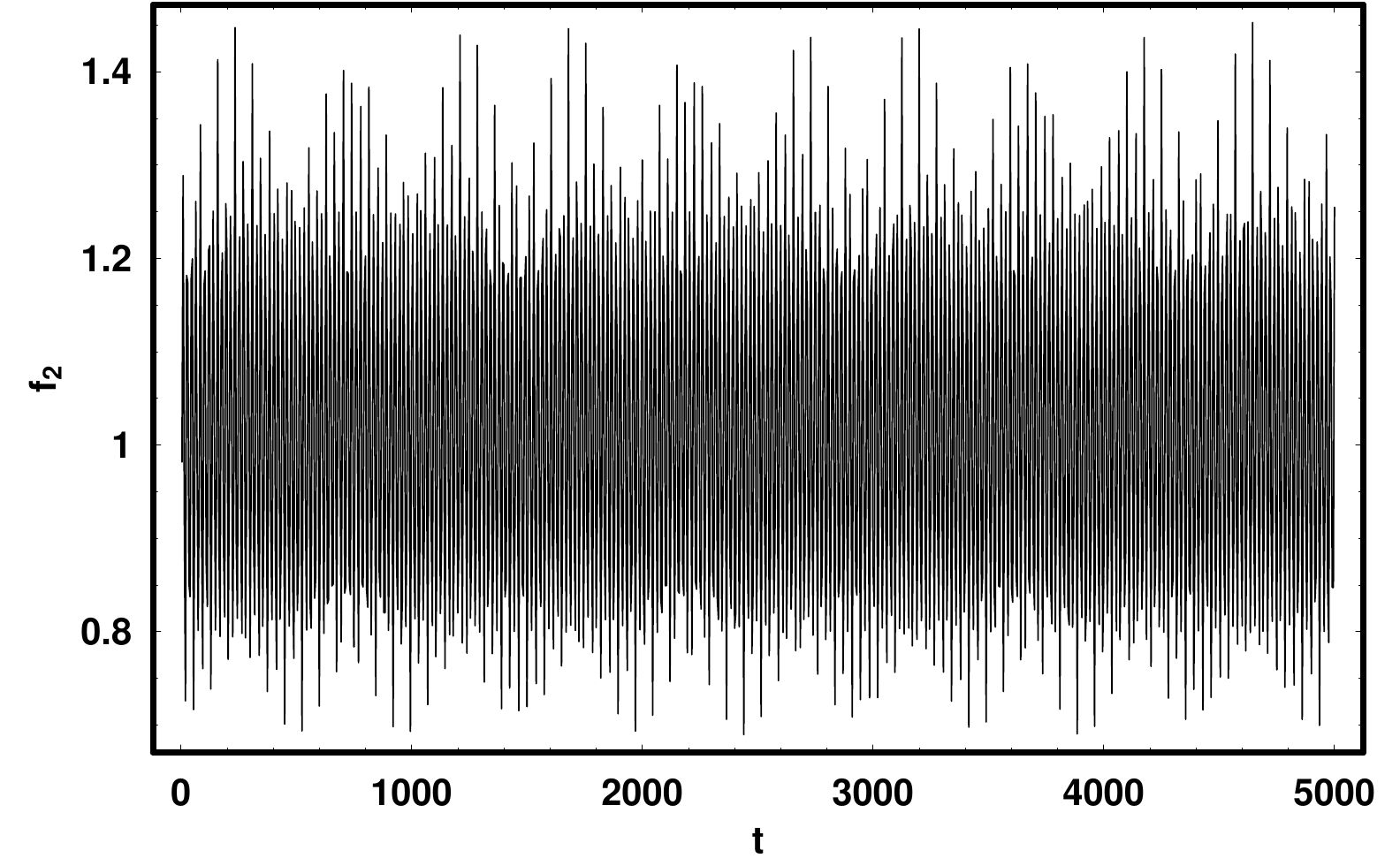}\label{Fig-5d}}
\caption{}
\centering
\hfill
\subfloat[ A quasi periodic orbit in the 2D global potential.]{\includegraphics[angle=0, scale=0.45]{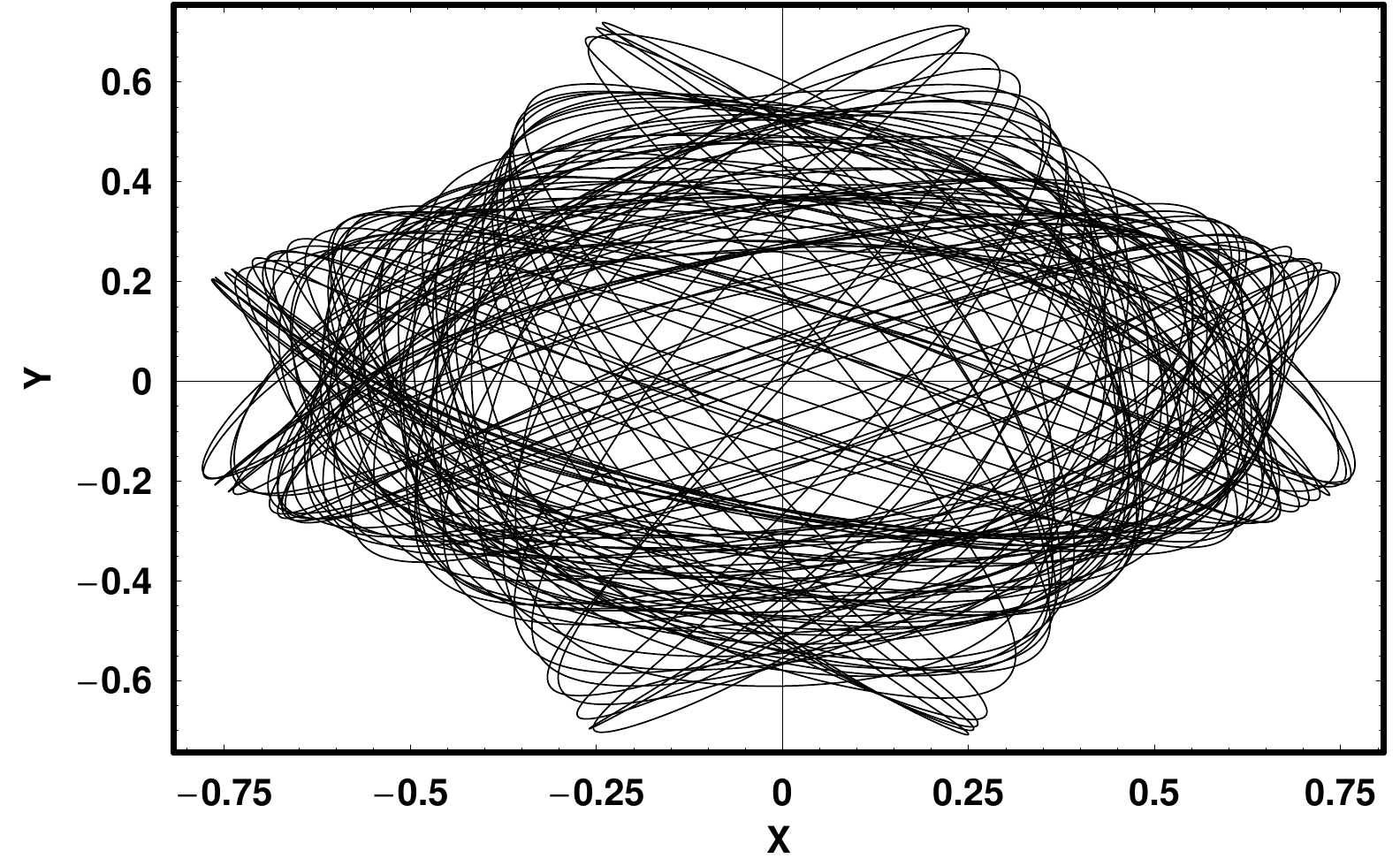}\label{Fig-6a}}\hfill
\subfloat[ The corresponding L.C.E.]{\includegraphics[angle=0, scale=0.45]{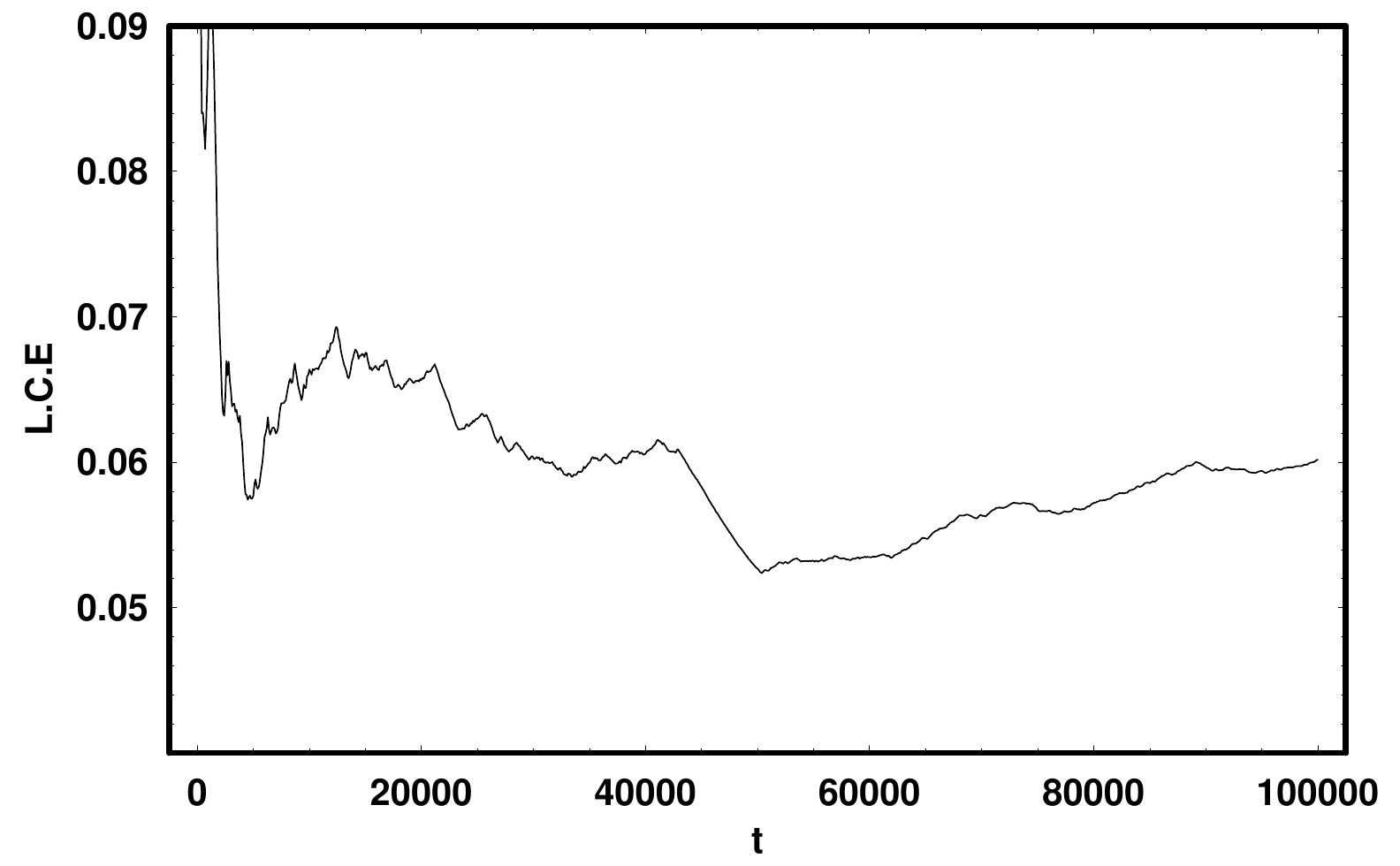}\label{Fig-6b}}\hfill\\
\hfill
\subfloat[ The $S(c)$ spectrum]{\includegraphics[angle=0, scale=0.45]{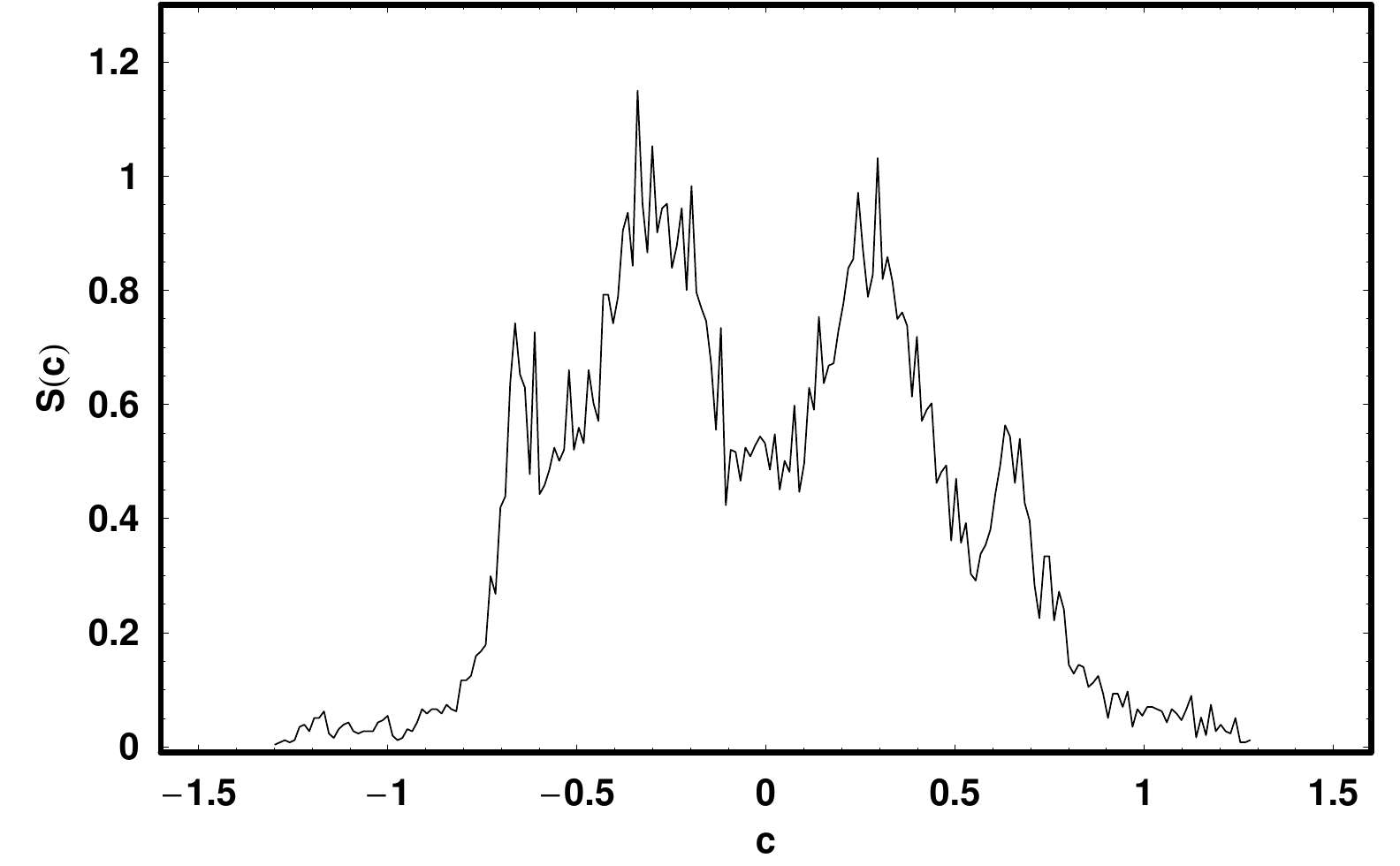}\label{Fig-6c}}\hfill
\subfloat[ The corresponding $f_2$-indicator.]{\includegraphics[angle=0, scale=0.45]{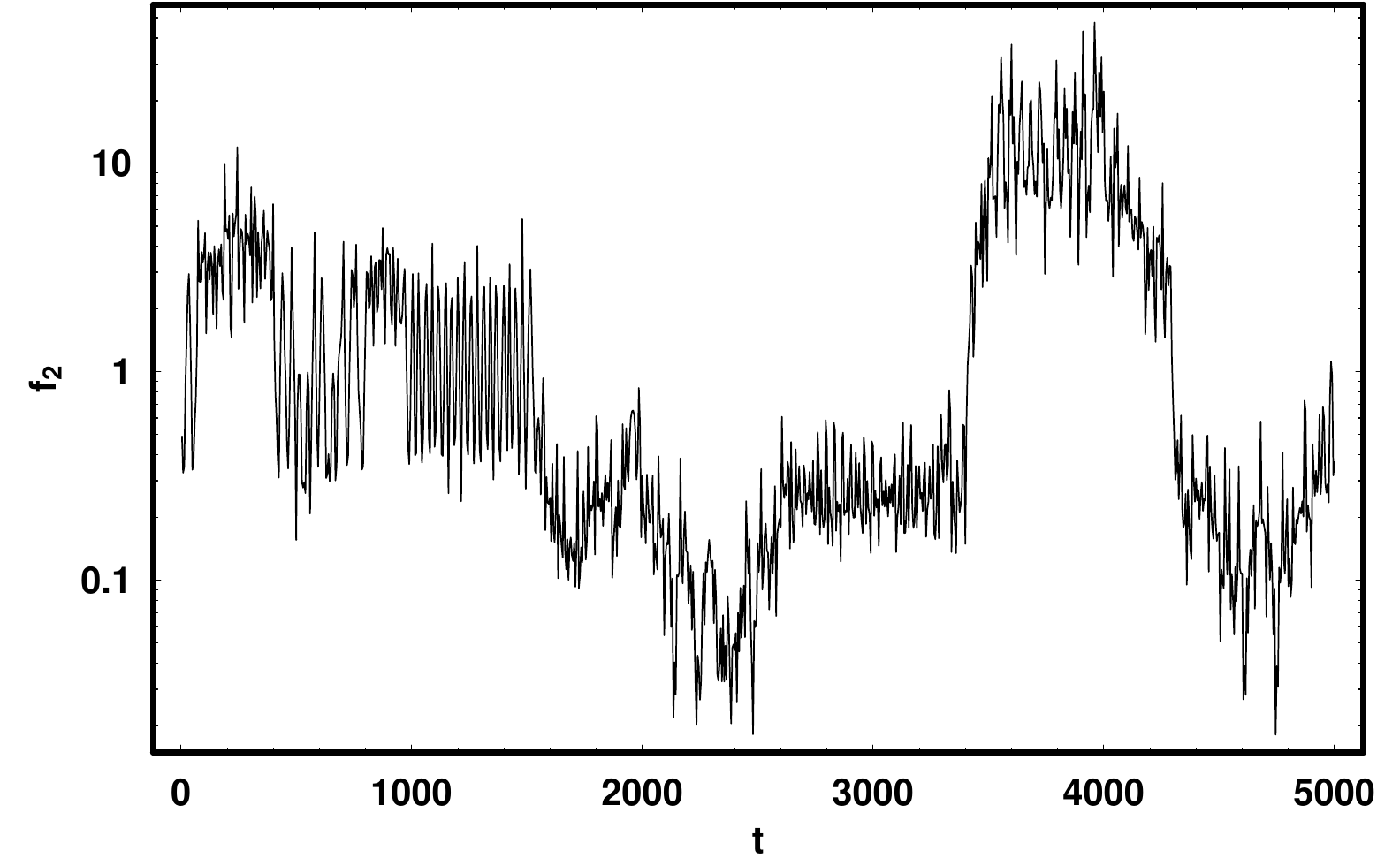}\label{Fig-6d}}\hfill
\caption{For a  chaotic orbit in the 2D local potential. See text for details.}
\end{figure}

\begin{figure}[htbp]
\centering
\subfloat[ A chaotic orbit in the 3D global potential.]{\includegraphics[angle=0, scale=0.30]{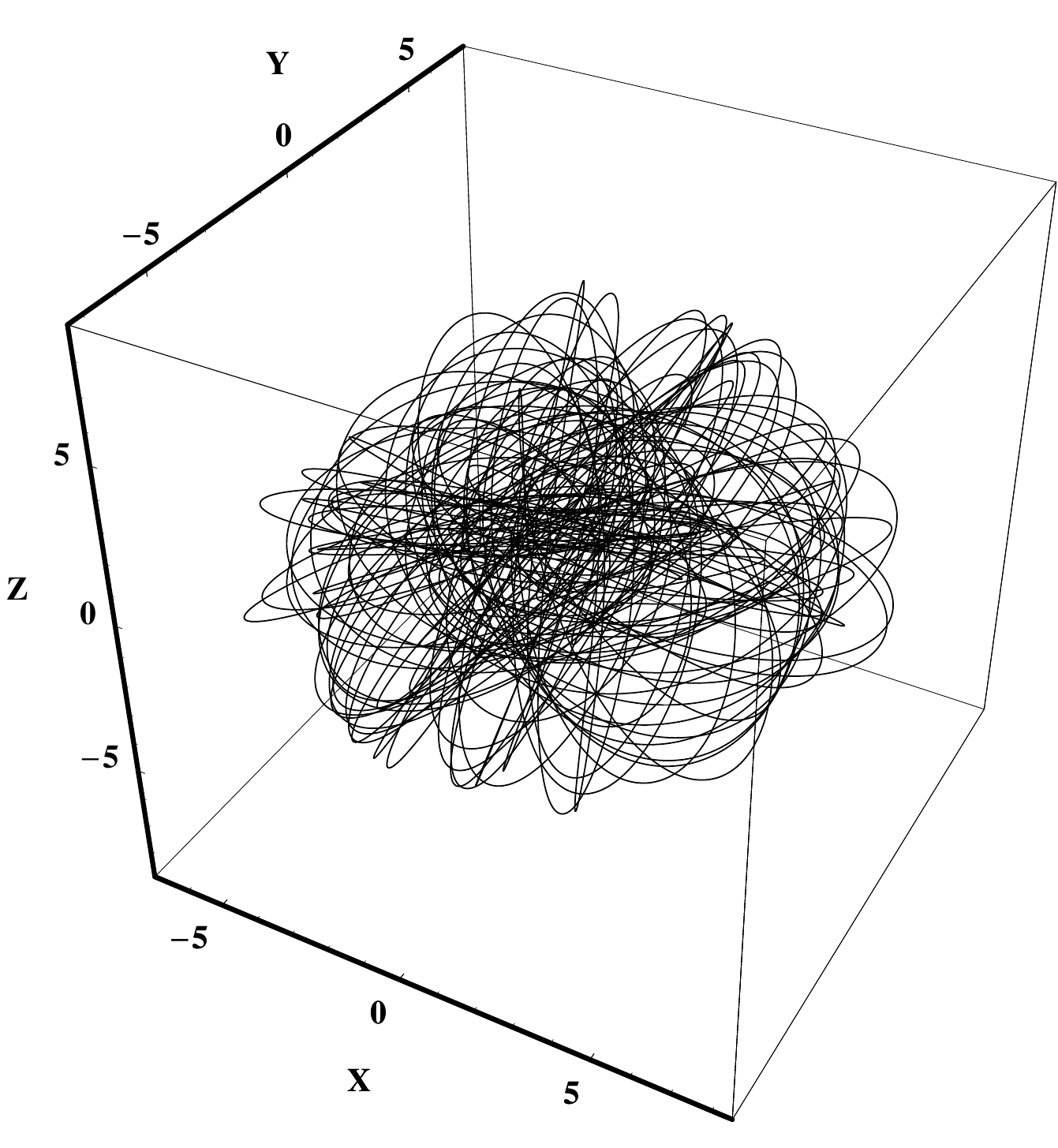}\label{Fig-7a}}
\hspace*{1cm}
\subfloat[ The corresponding L.C.E.]{\includegraphics[angle=0, scale=0.45]{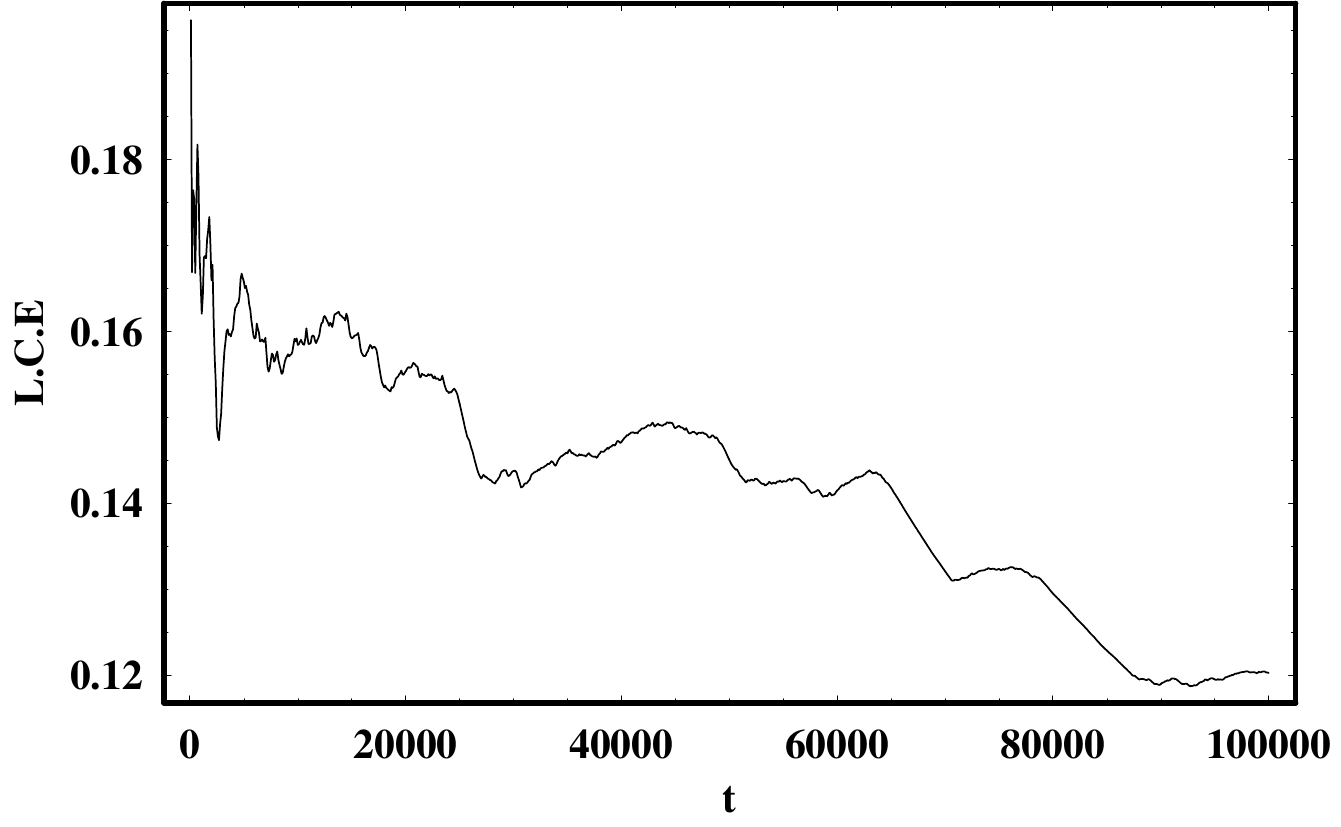}\label{Fig-7b}}\\
\subfloat[ The $S(c)$ spectrum]{\includegraphics[angle=0, scale=0.45]{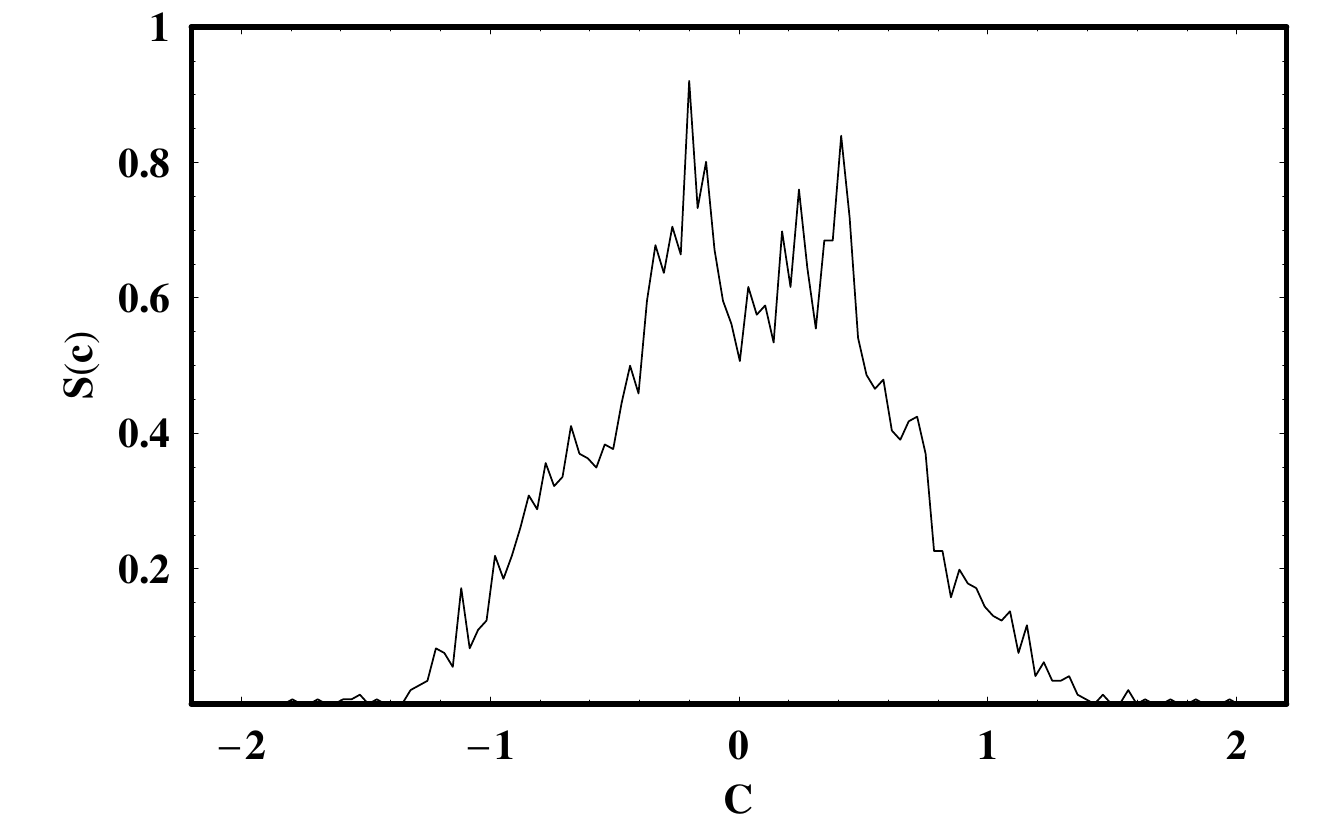}\label{Fig-7c}}
\hspace*{1cm}
\subfloat[ The corresponding $f_3$-indicator.]{\includegraphics[angle=0, scale=0.45]{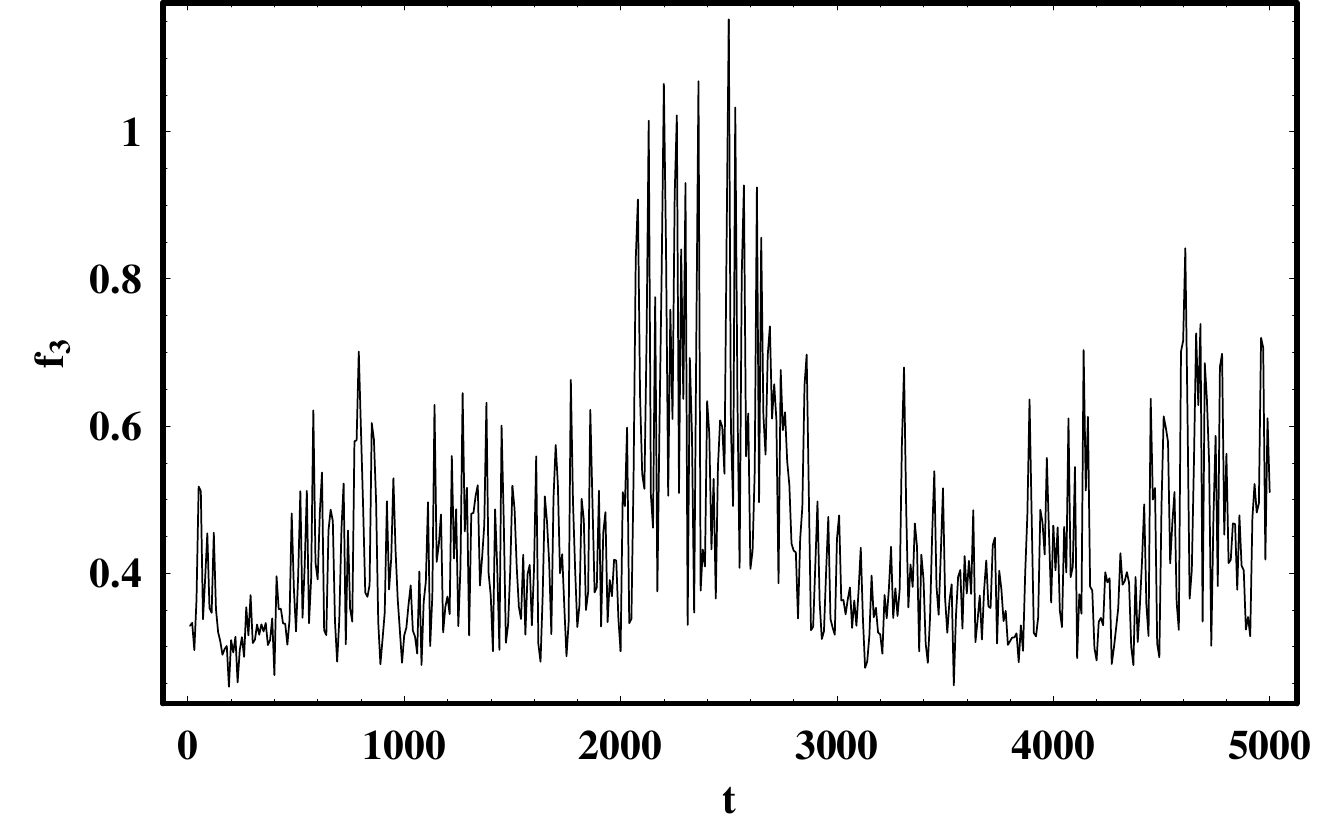}\label{Fig-7d}}
\caption{}
\centering
\subfloat[ A chaotic orbit in the 3D global potential.]{\includegraphics[angle=0, scale=0.30]{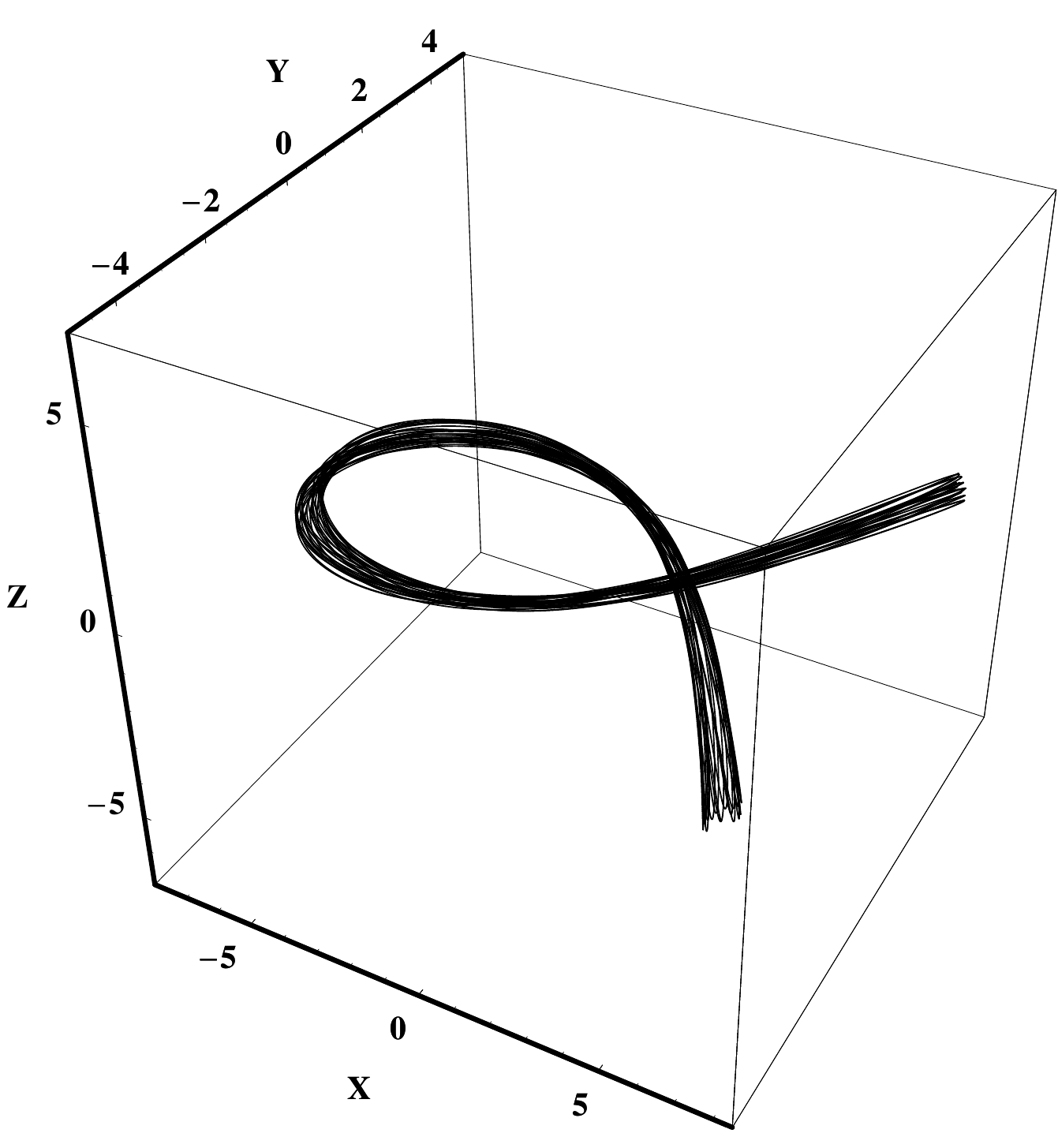}\label{Fig-8a}}
\hspace*{1cm}
\subfloat[ The corresponding L.C.E.]{\includegraphics[angle=0, scale=0.45]{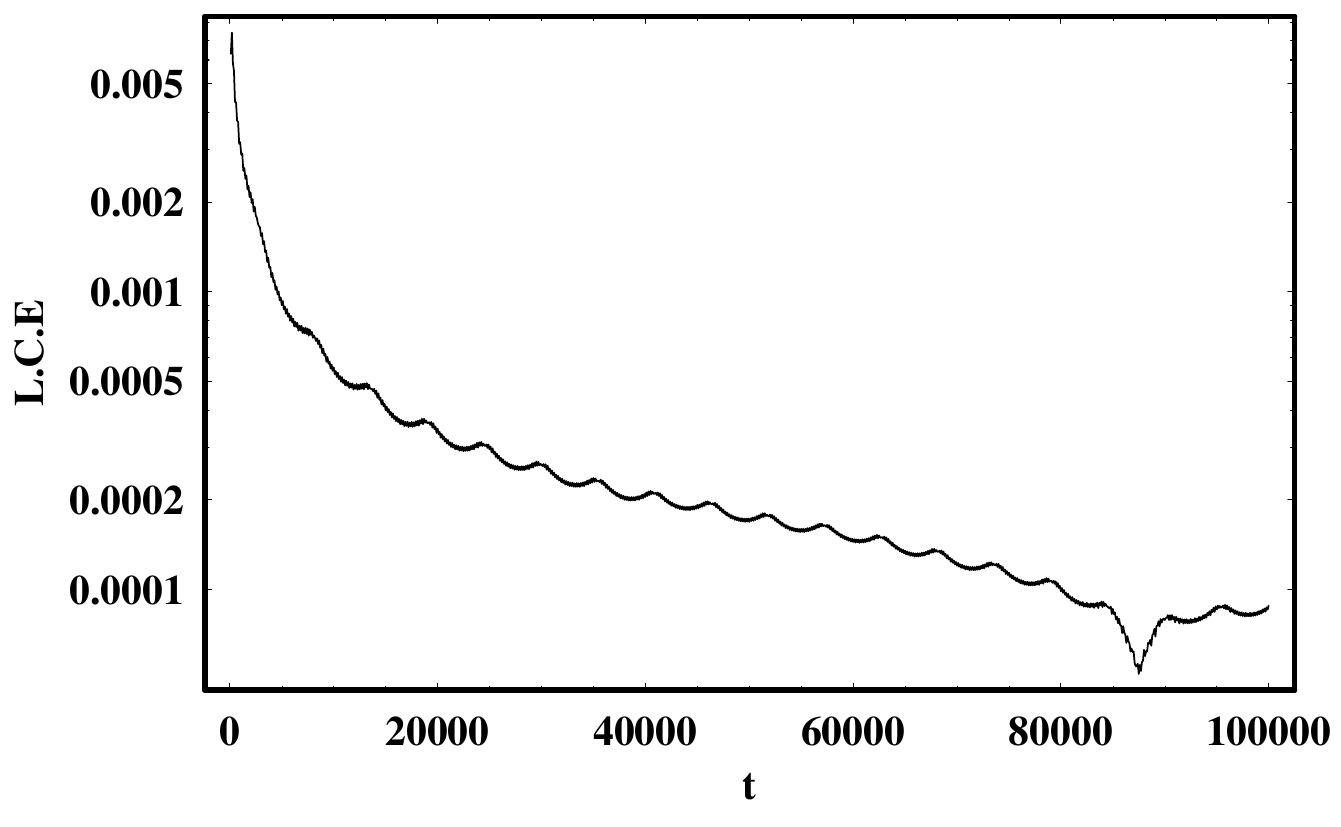}\label{Fig-8b}}\\
\subfloat[ The $S(c)$ spectrum]{\includegraphics[angle=0, scale=0.45]{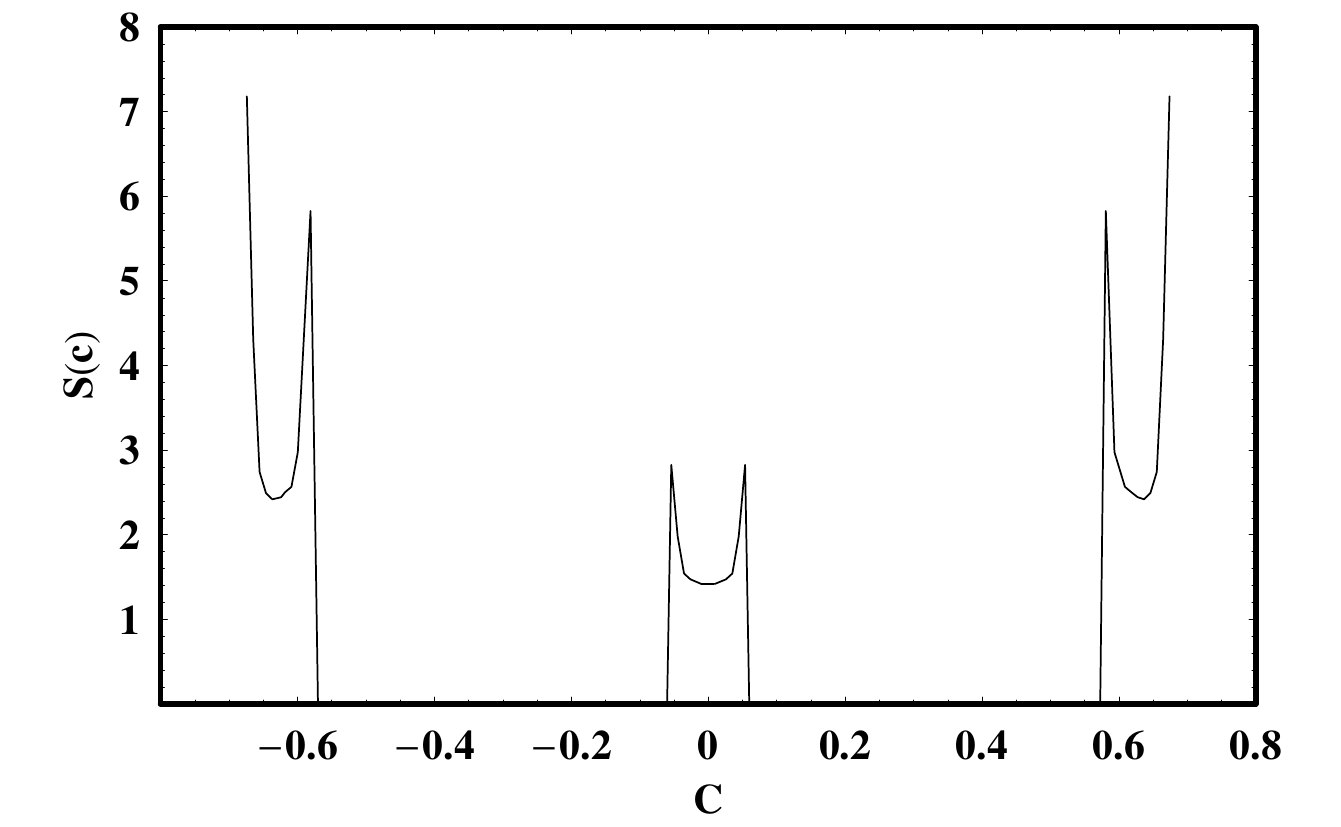}\label{Fig-8c}}
\hspace*{1cm}
\subfloat[ The corresponding $f_3$-indicator.]{\includegraphics[angle=0, scale=0.45]{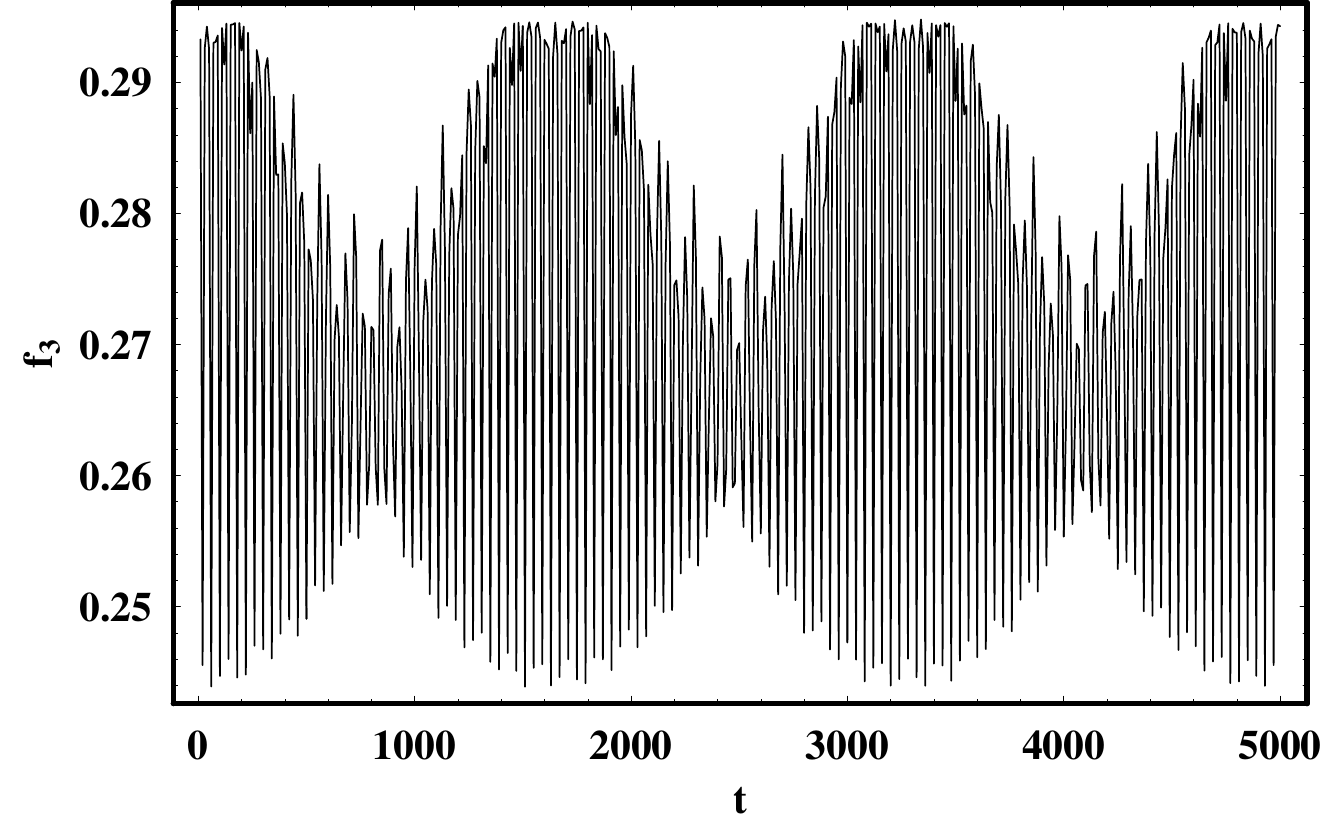}\label{Fig-8d}}
\caption{For a regular resonant orbit in the 3D global potential. See text for details.}
\end{figure}

\begin{figure}[htbp]
\centering
\subfloat[ A chaotic orbit in the 3D global potential.]{\includegraphics[angle=0, scale=0.35]{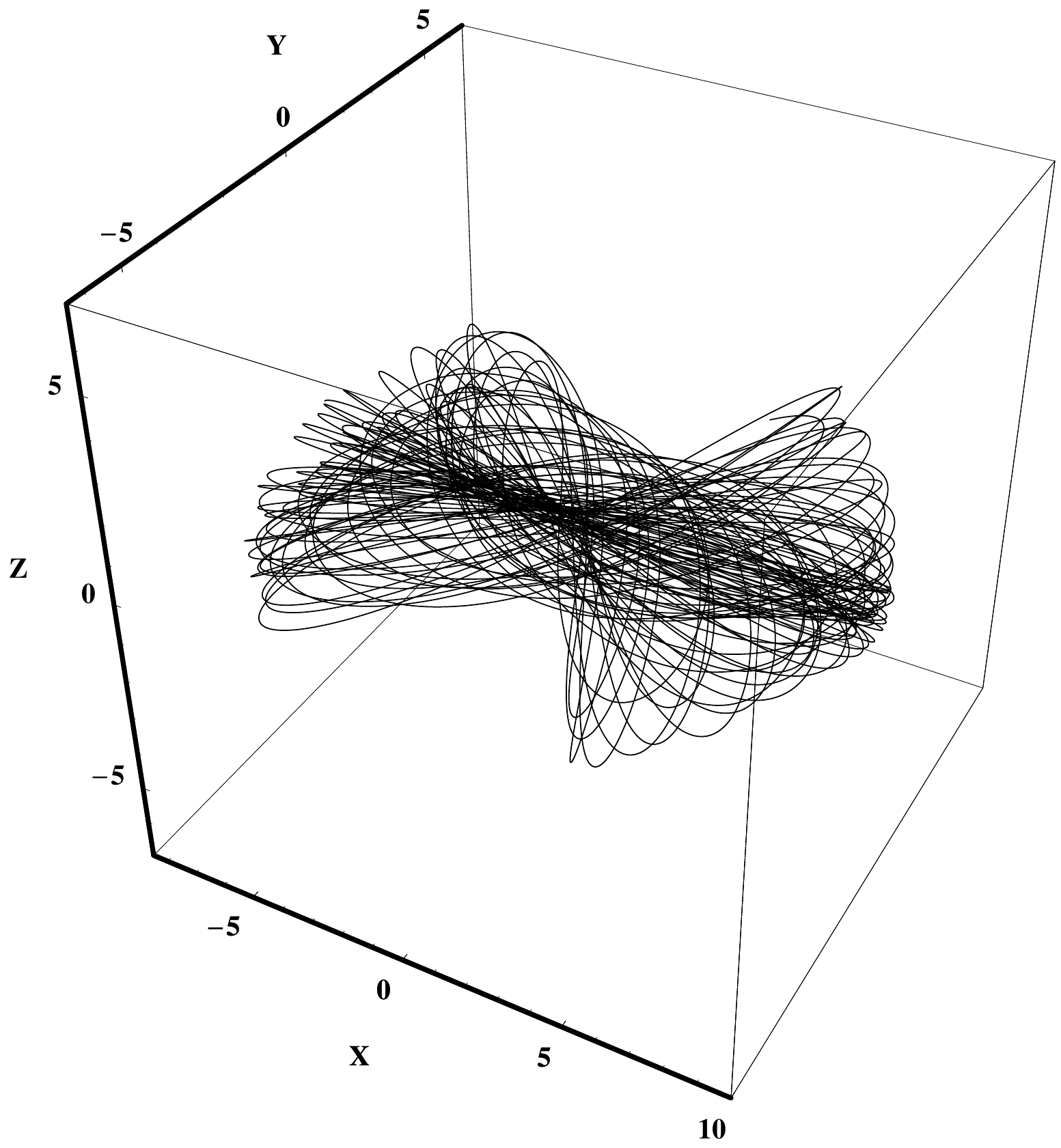}\label{Fig-9a}}
\hspace*{1cm}
\subfloat[ The corresponding L.C.E.]{\includegraphics[angle=0, scale=0.45]{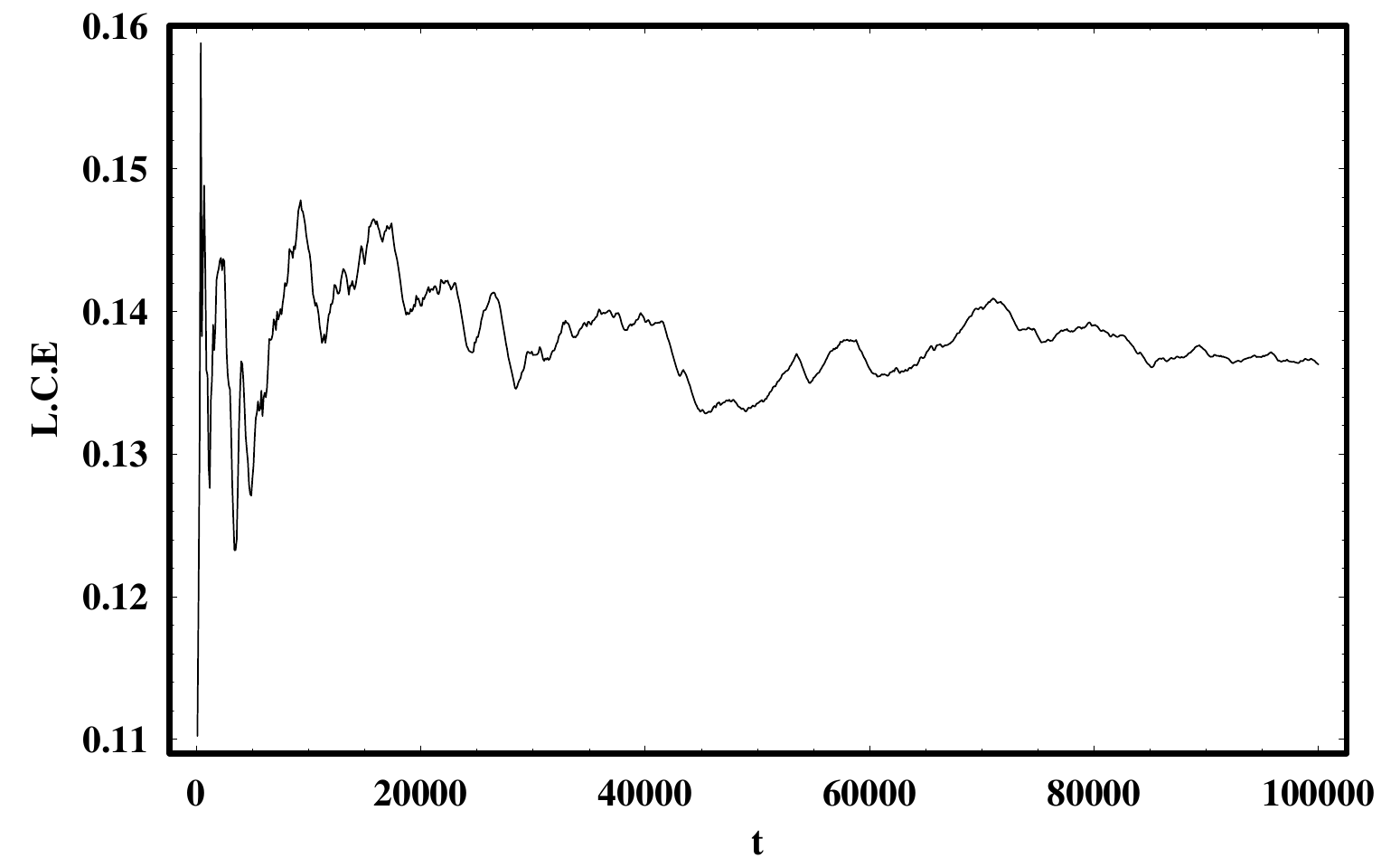}\label{Fig-9b}}\\
\subfloat[ The $S(c)$ spectrum]{\includegraphics[angle=0, scale=0.45]{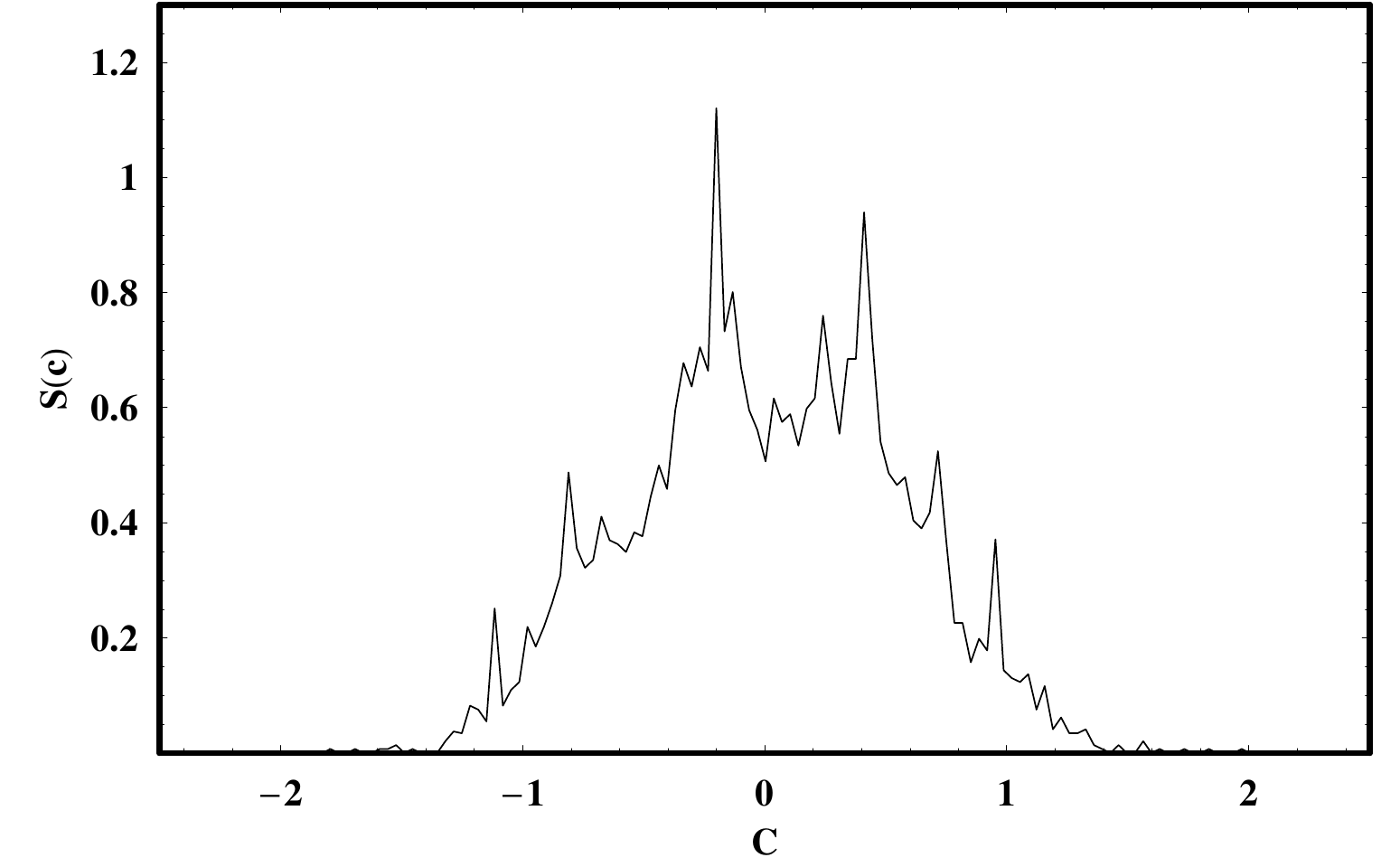}\label{Fig-9c}}
\hspace*{1cm}
\subfloat[ The corresponding $f_3$-indicator.]{\includegraphics[angle=0, scale=0.45]{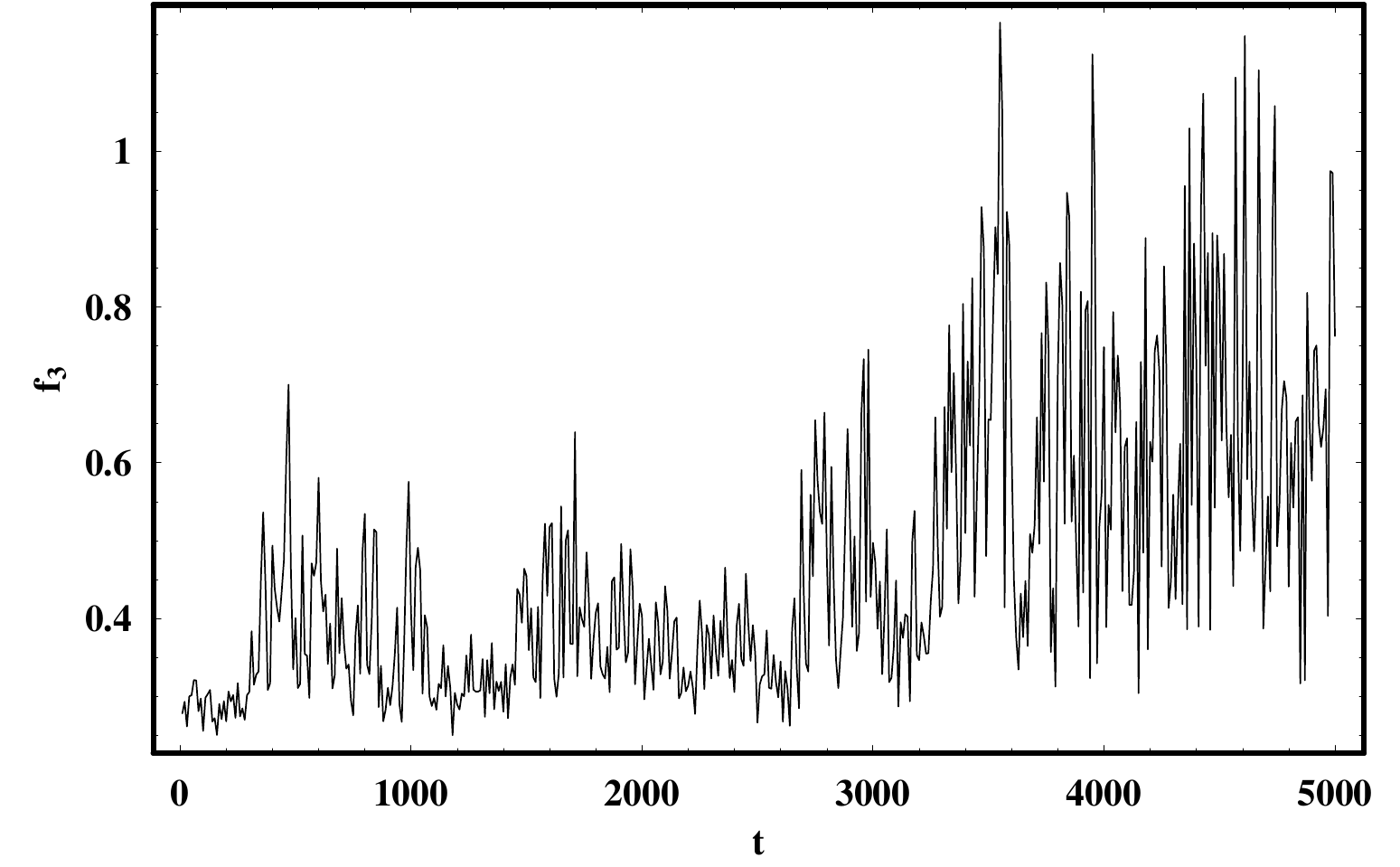}\label{Fig-9d}}
\caption{When $z_0=0.6$. See text for details..}
\end{figure}

\begin{figure}[htbp]
\centering
\subfloat[ A chaotic orbit in the 3D local potential.]{\includegraphics[angle=0, scale=0.30]{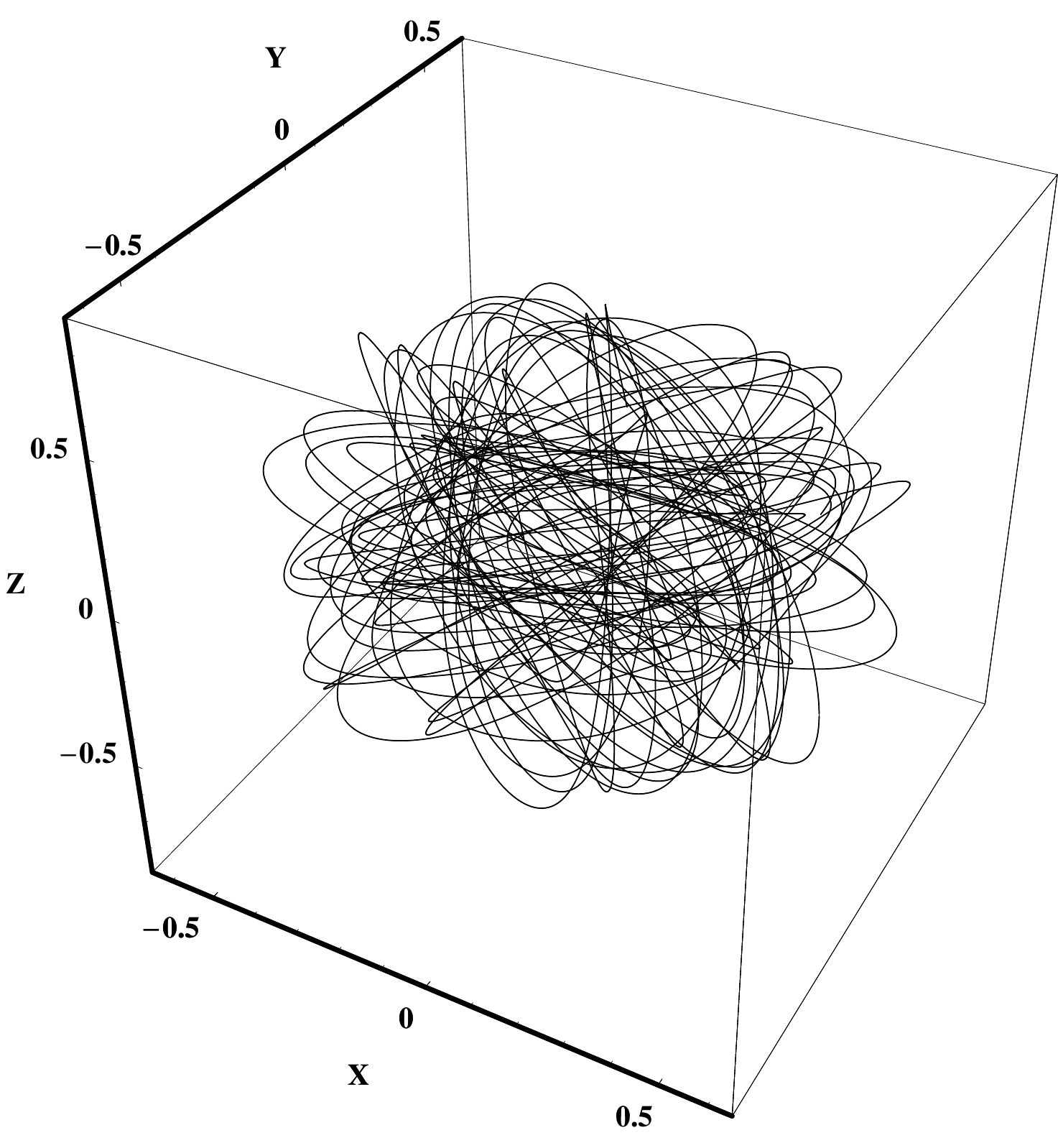}\label{Fig-10a}}
\hspace*{1cm}
\subfloat[ The corresponding L.C.E.]{\includegraphics[angle=0, scale=0.45]{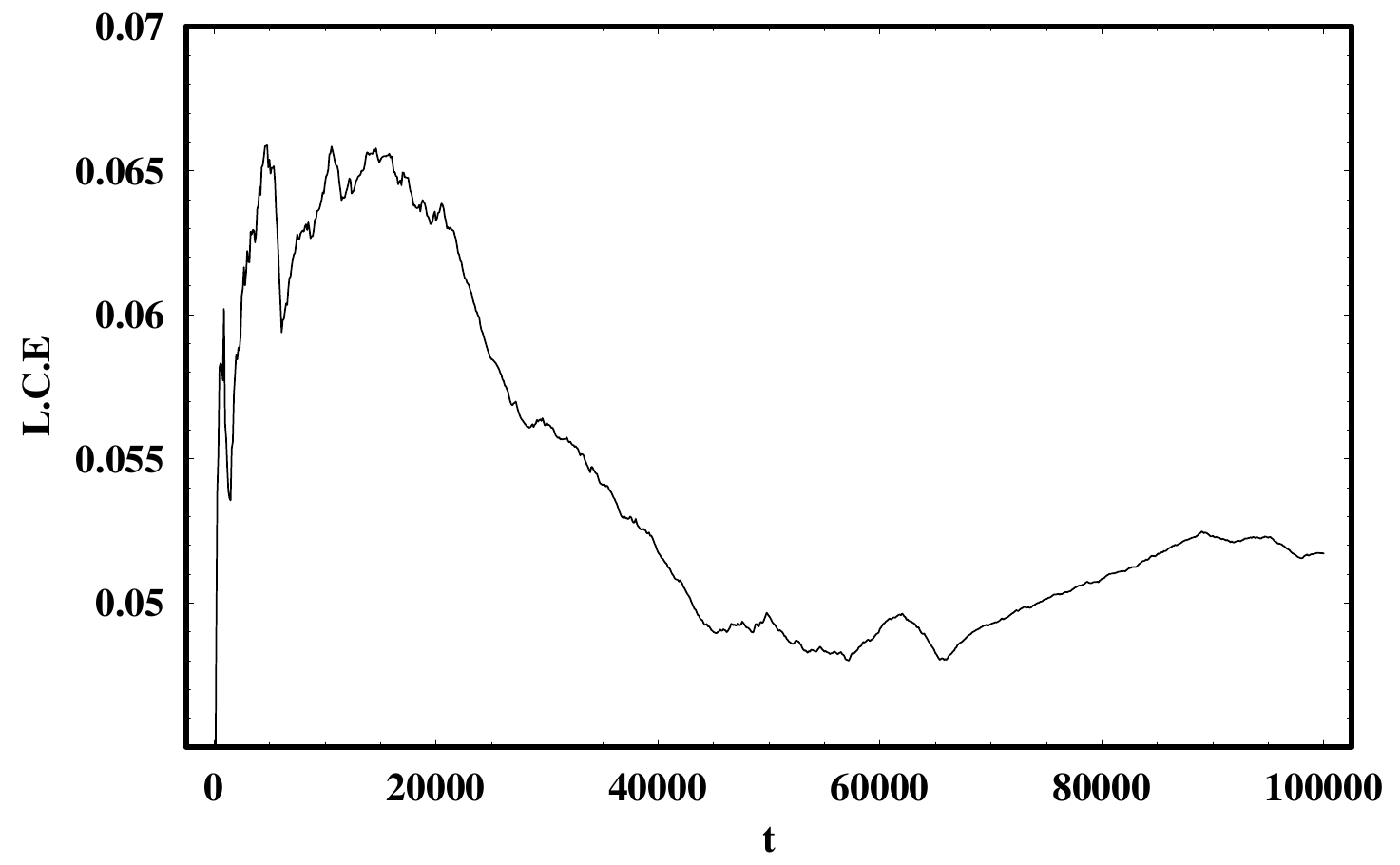}\label{Fig-10b}}\\
\subfloat[ The $S(c)$ spectrum]{\includegraphics[angle=0, scale=0.45]{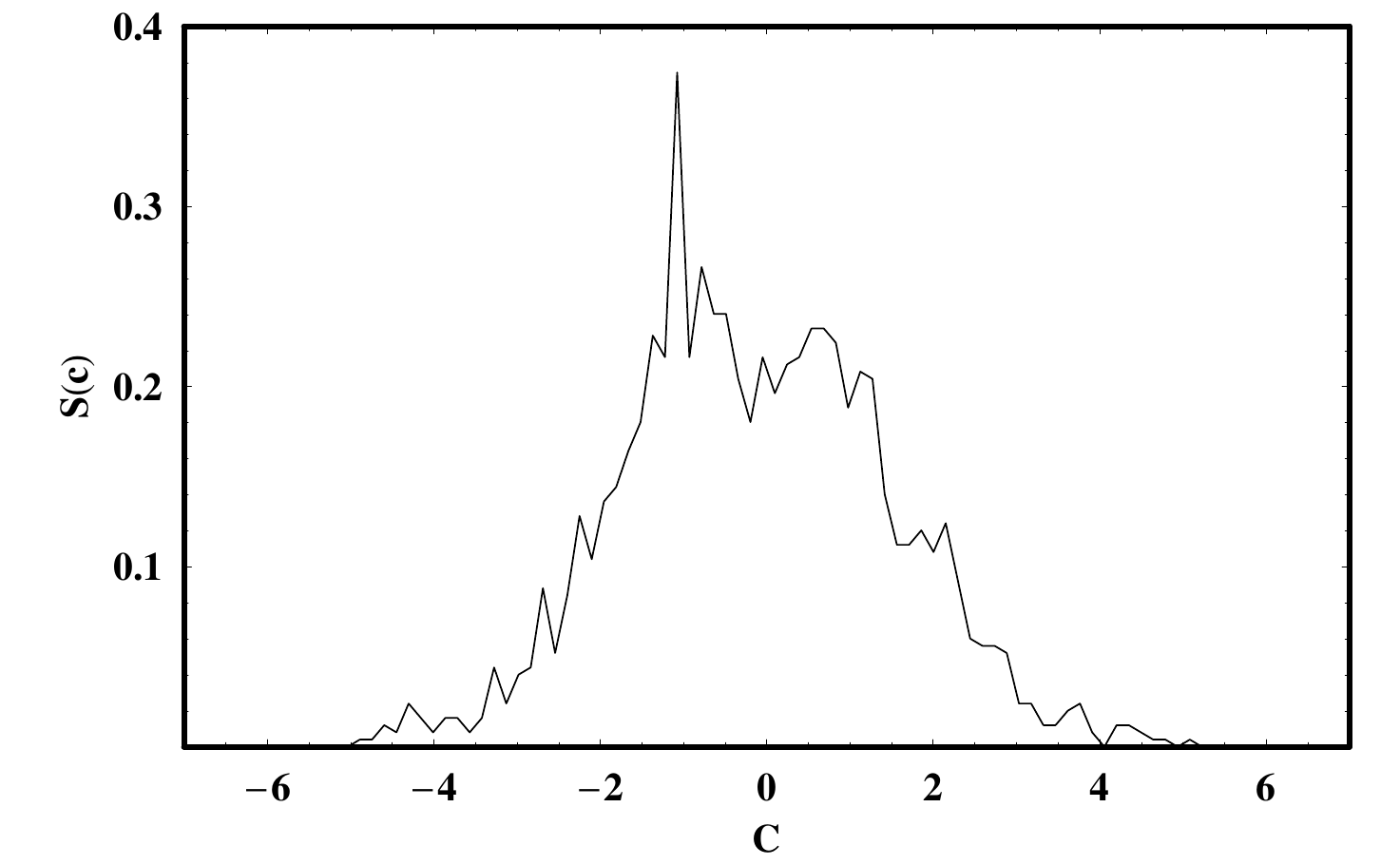}\label{Fig-10c}}
\hspace*{1cm}
\subfloat[ The corresponding $f_3$-indicator.]{\includegraphics[angle=0, scale=0.45]{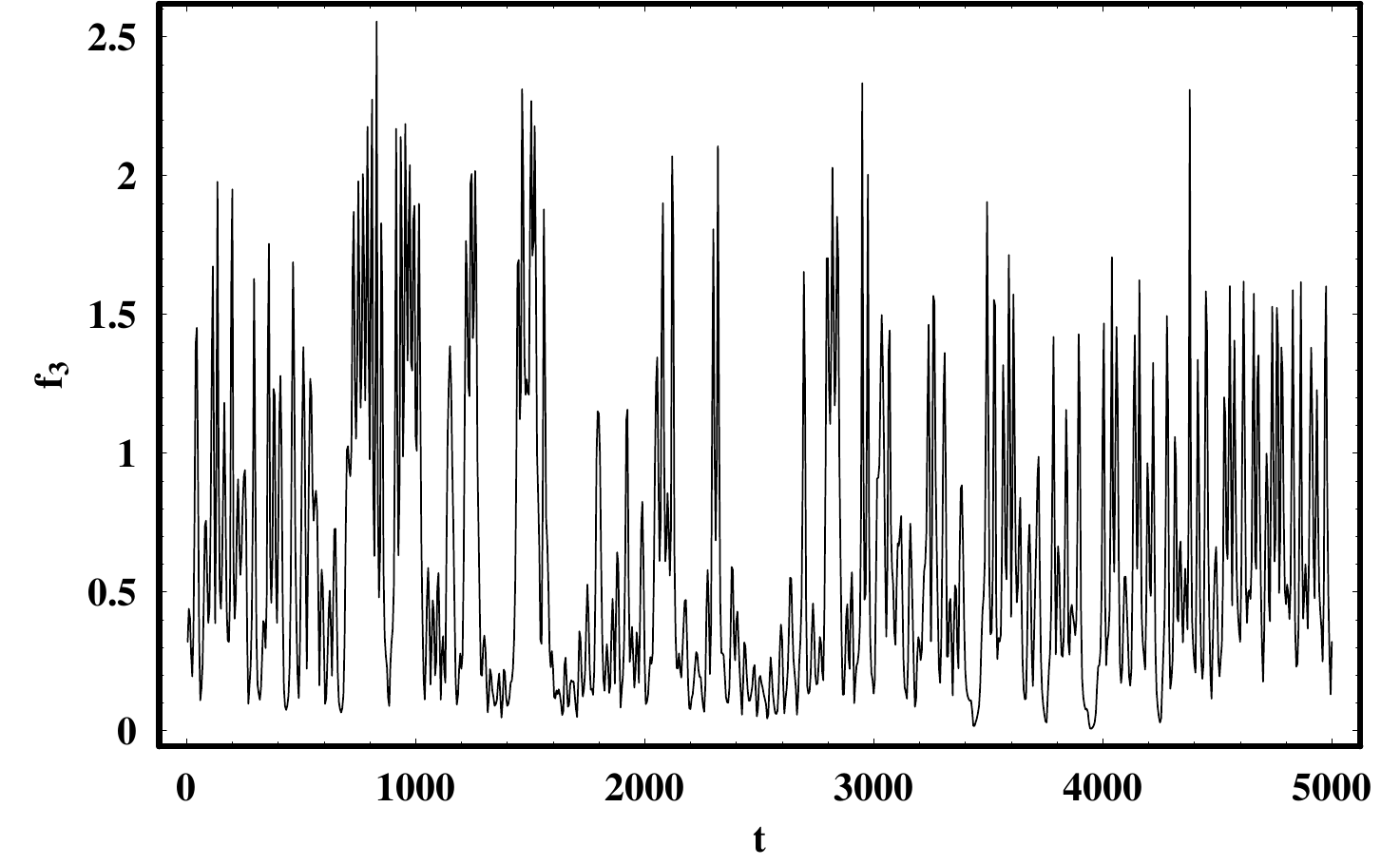}\label{Fig-10d}}
\caption{}
\end{figure}

\begin{figure}[htbp]
\centering
\subfloat[ A chaotic orbit in the 3D local potential.]{\includegraphics[angle=0, scale=0.30]{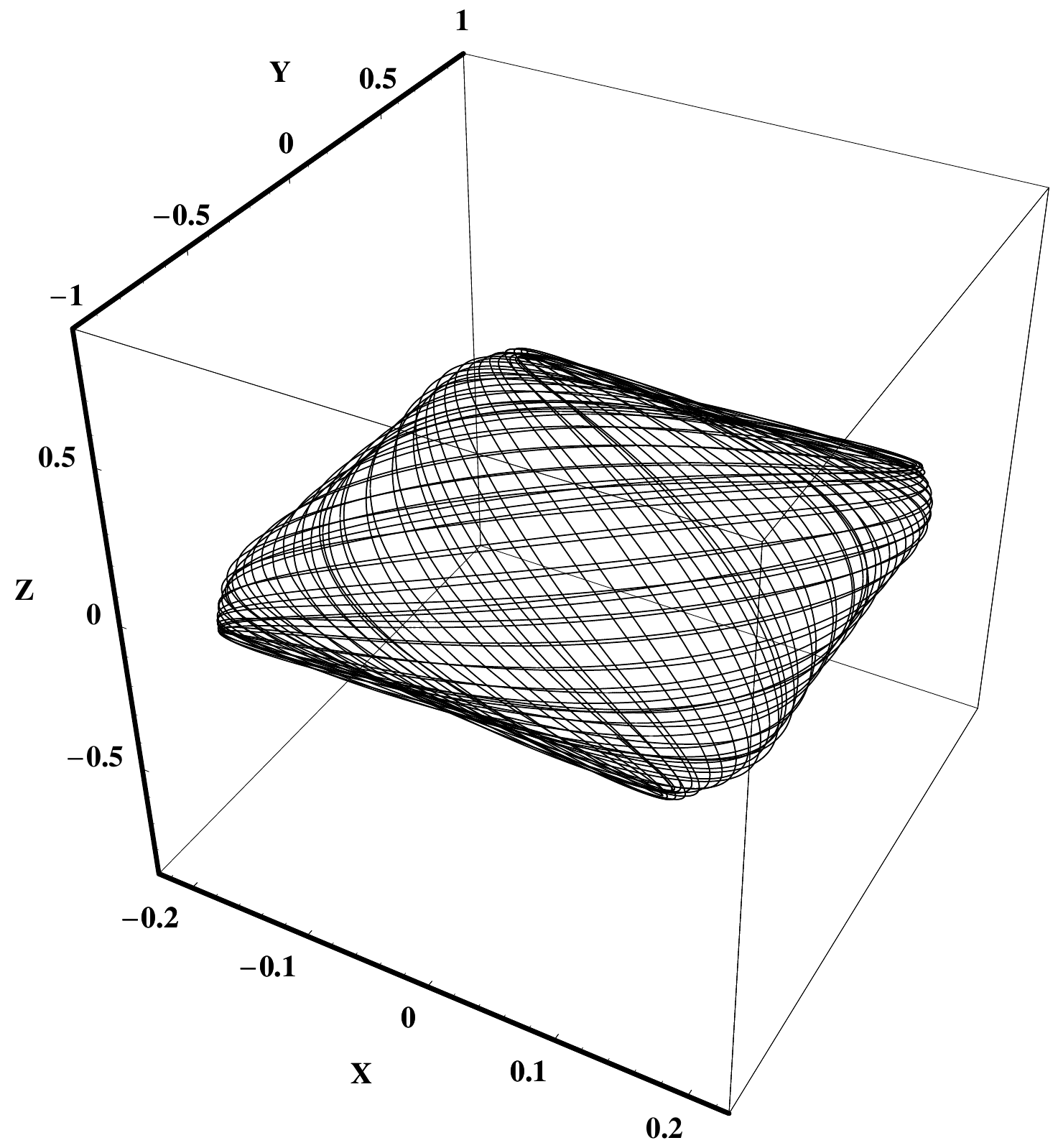}\label{Fig-11a}}
\hspace*{1cm}
\subfloat[ The corresponding L.C.E.]{\includegraphics[angle=0, scale=0.45]{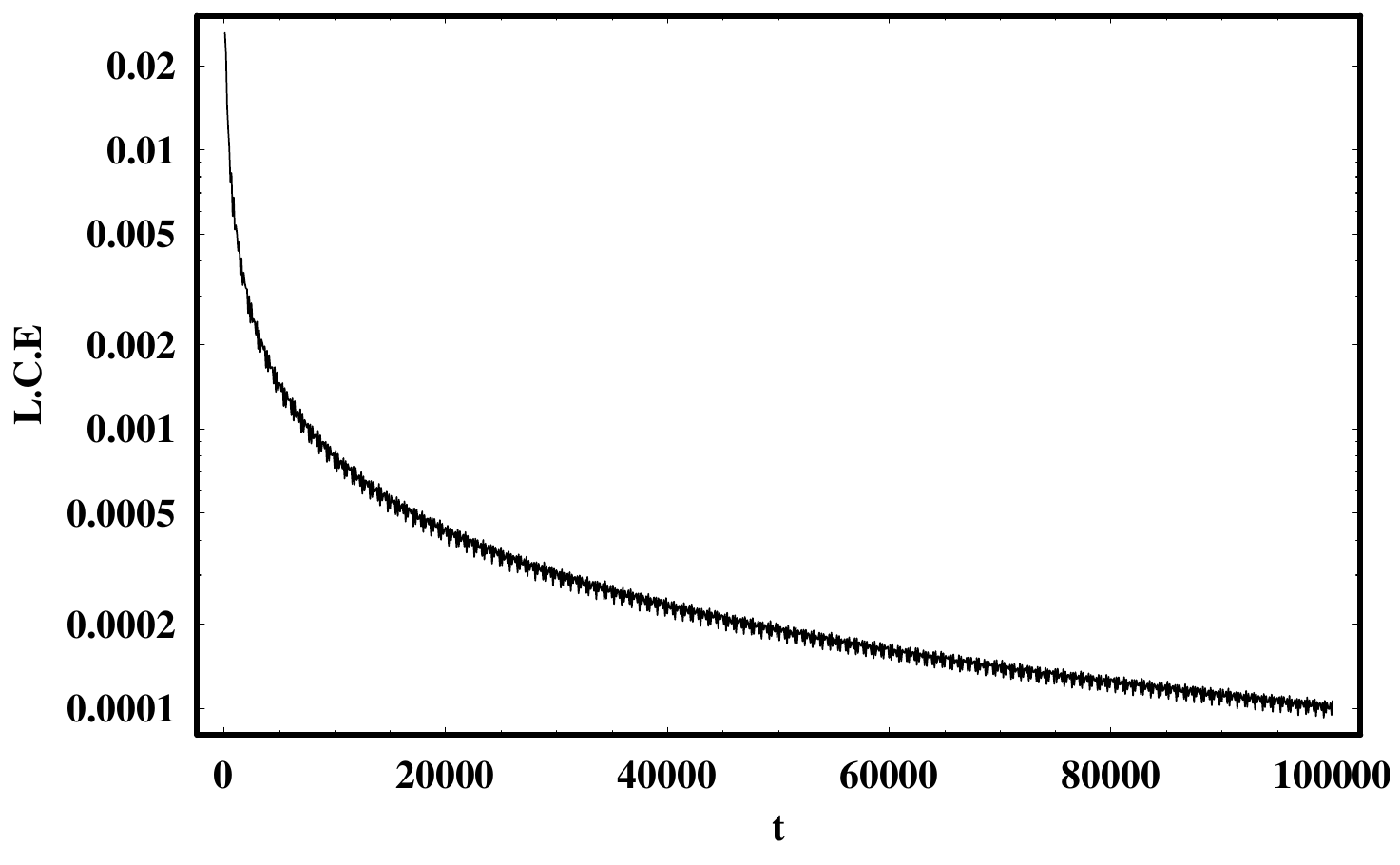}\label{Fig-11b}}\\
\subfloat[ The $S(c)$ spectrum]{\includegraphics[angle=0, scale=0.45]{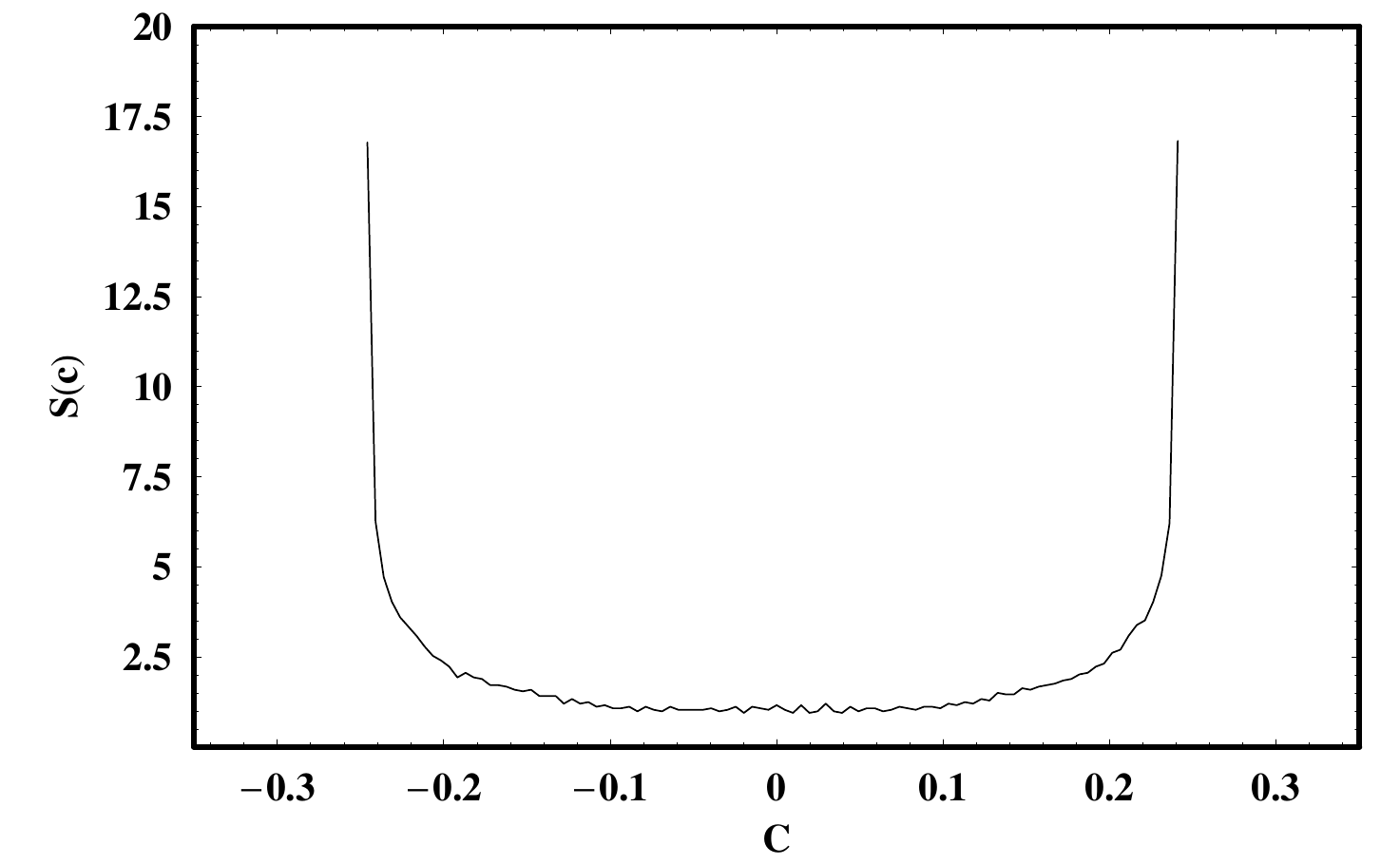}\label{Fig-11c}}
\hspace*{1cm}
\subfloat[ The corresponding $f_3$-indicator.]{\includegraphics[angle=0, scale=0.45]{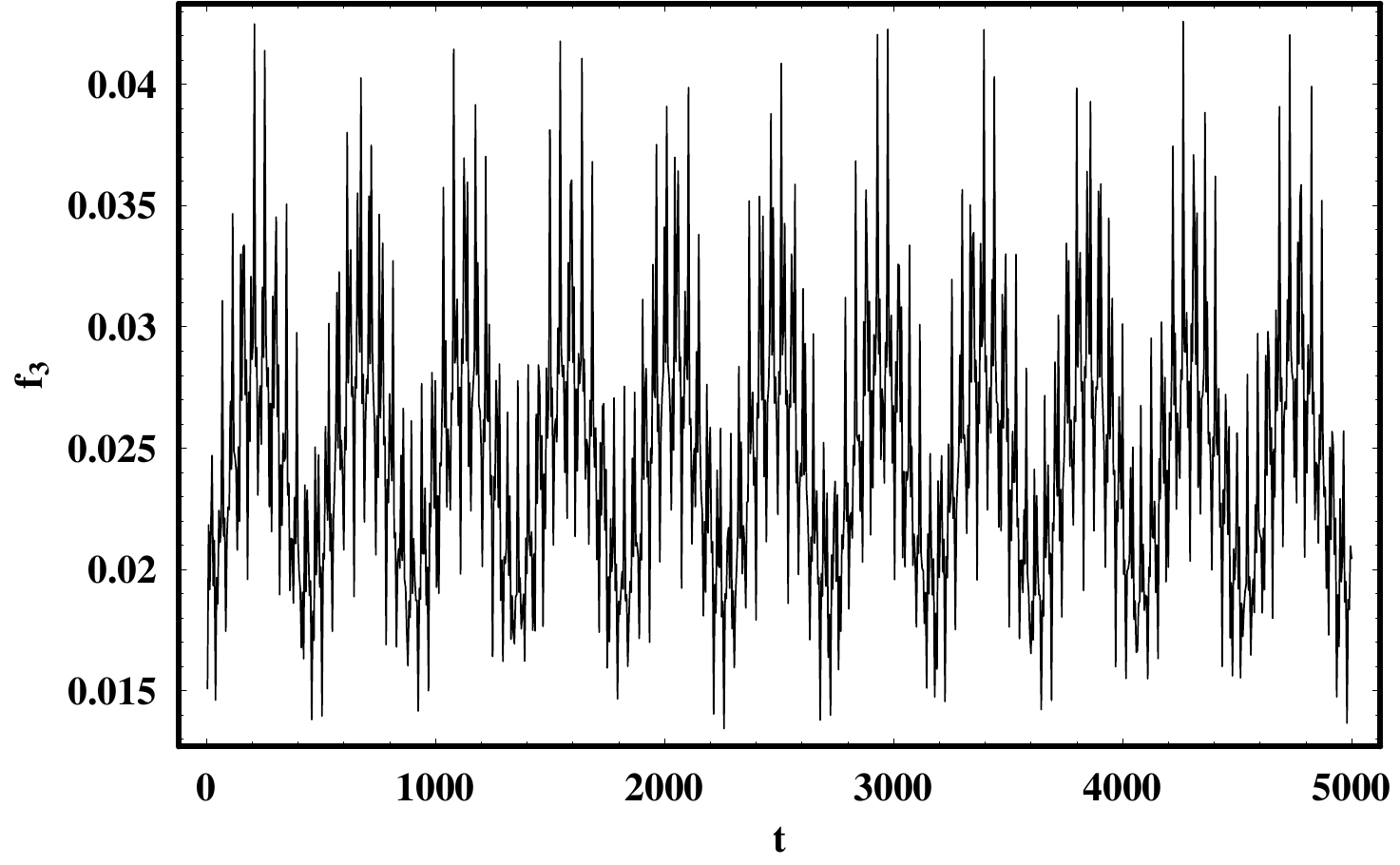}\label{Fig-11d}}
\caption{For a regular orbit of the 3D local potential. See text for details.}
\end{figure}

%\newpage
\begin{figure}[htbp]
\centering
\subfloat[ A chaotic orbit in the 3D local potential.]{\includegraphics[angle=0, scale=0.30]{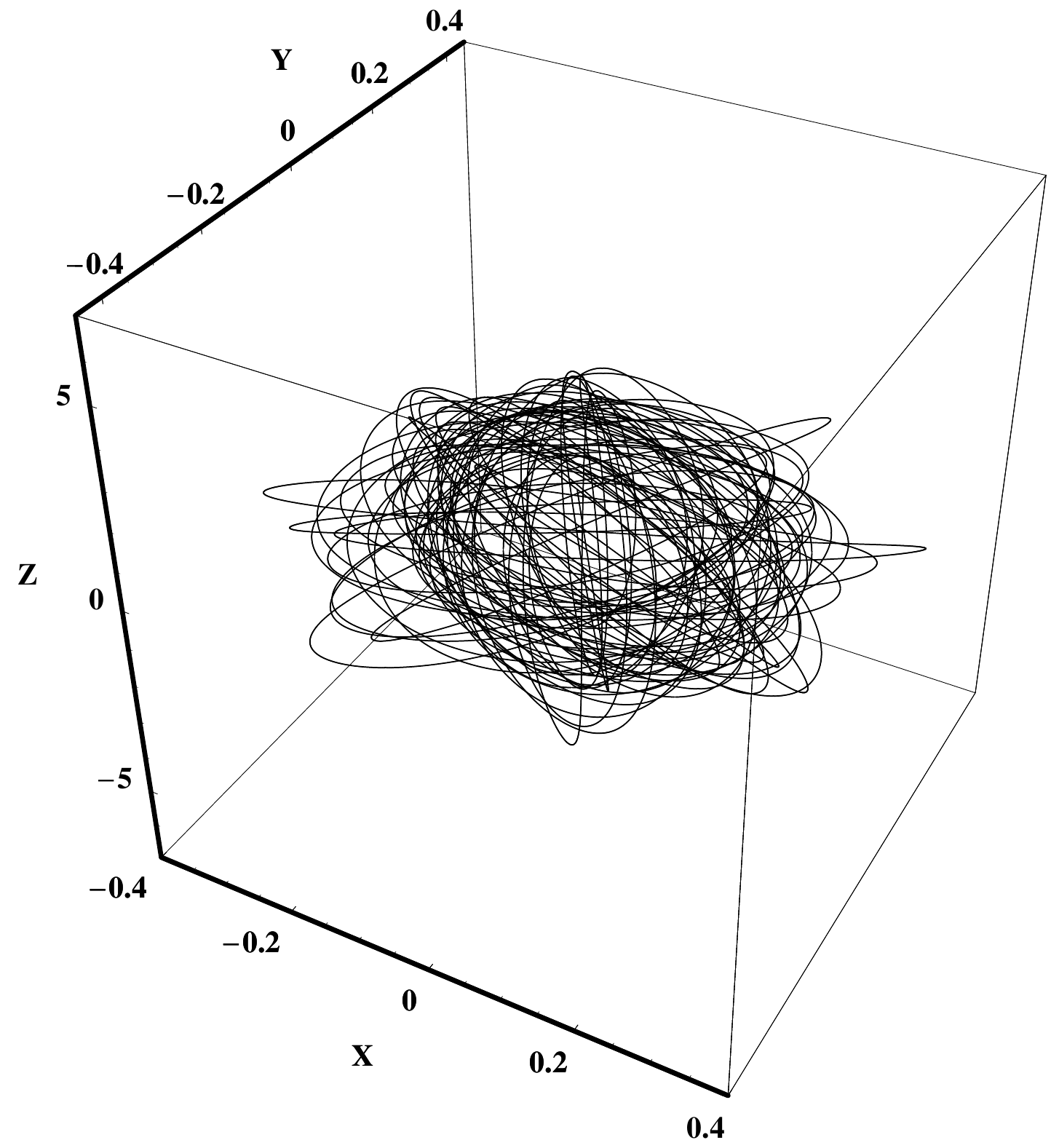}\label{Fig-12a}}
\hspace*{1cm}
\subfloat[ The corresponding L.C.E.]{\includegraphics[angle=0, scale=0.45]{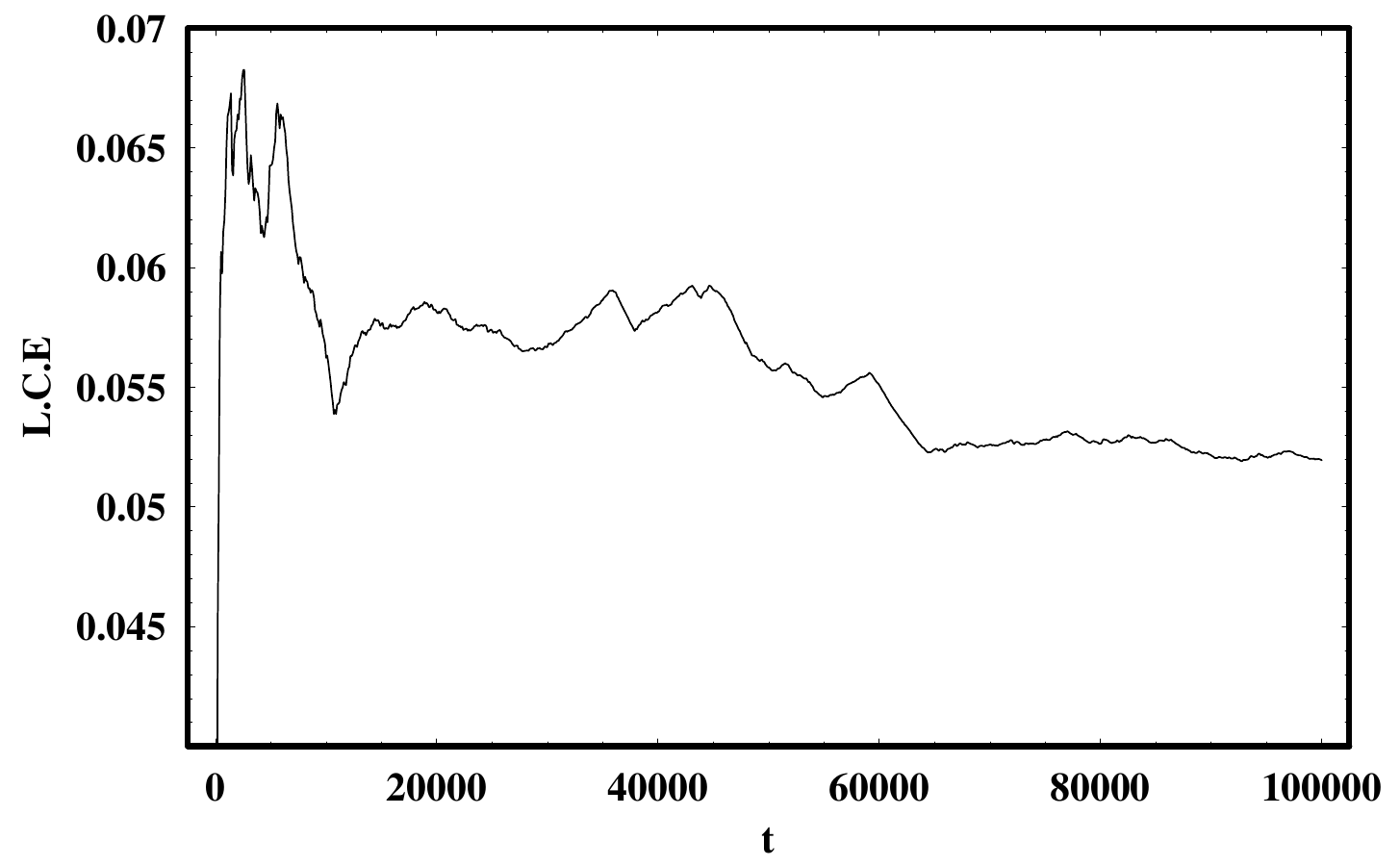}\label{Fig-12b}}\\
\subfloat[ The $S(c)$ spectrum]{\includegraphics[angle=0, scale=0.45]{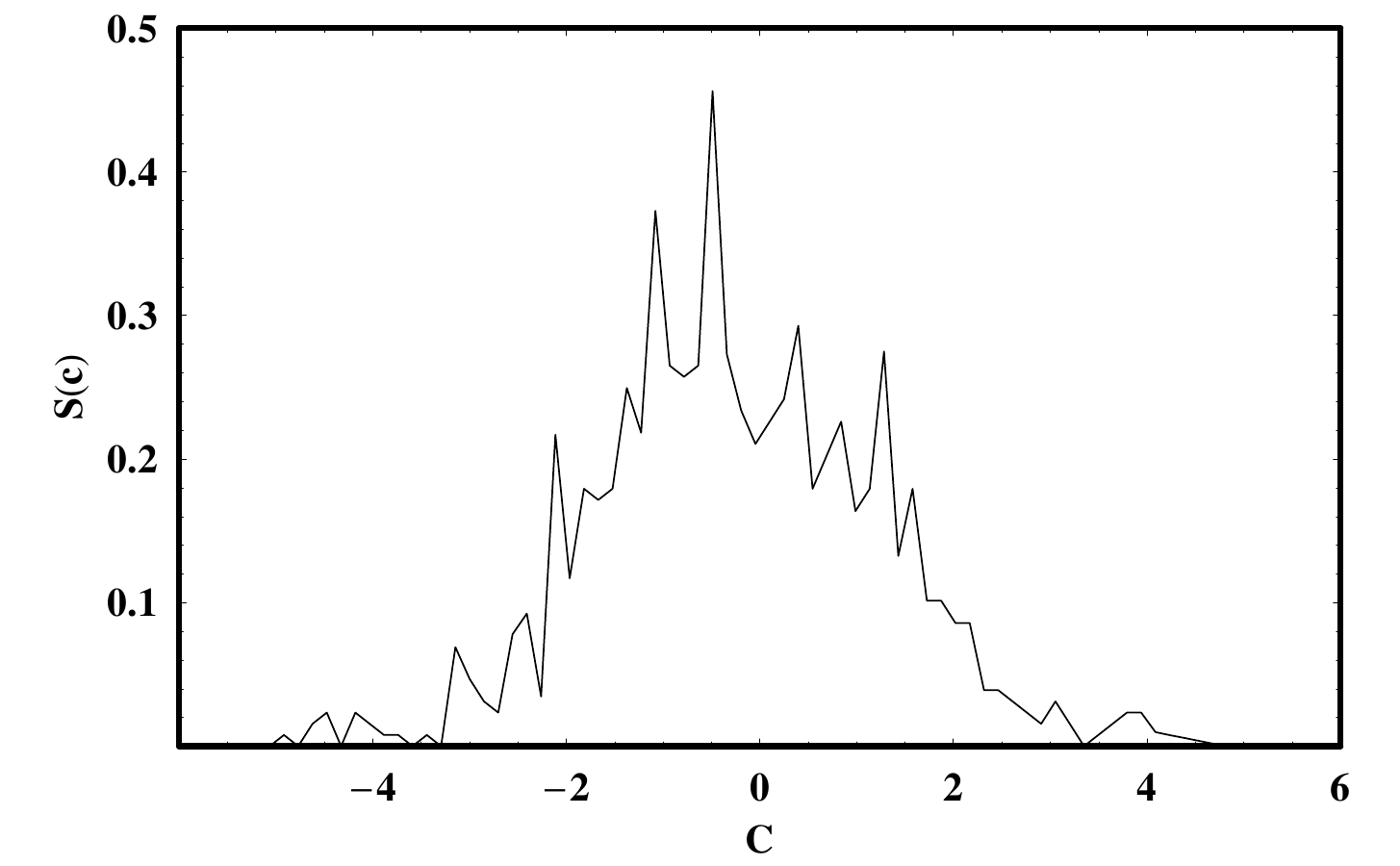}\label{Fig-12c}}
\hspace*{1cm}
\subfloat[ The corresponding $f_3$-indicator.]{\includegraphics[angle=0, scale=0.45]{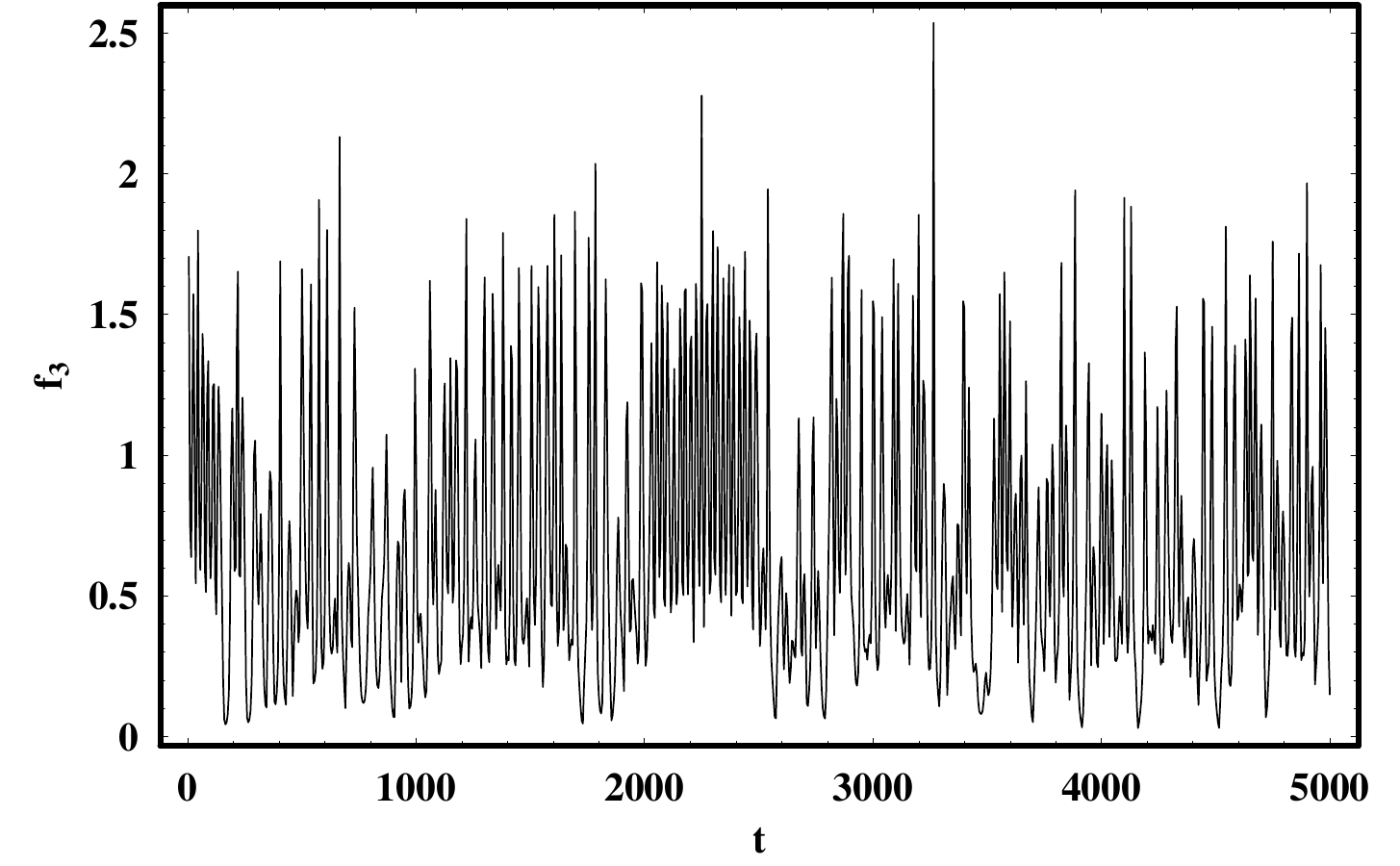}\label{Fig-12d}}
\caption{When $z_0=0.6$. See text for details.}
\end{figure}

Quite different are the results shown in Figure 4a-d. The orbit shown in Fig. 4a is the orbit producing the separatrix shown in the outer part of Fig. 1a. The initial conditions are: $x_0=-0.51, p_{x0}=23$. The values of all other parameters are as in Fig. 1a. The L.C.E in Fig. 4b has a value of about 0.017, while the $S(c)$ spectrum shown in Fig. 4c is asymmetric with a number of small and large peaks, which is also an indication of chaotic motion. Finally, the $f_2$-indicator in Fig. 4d appears to be highly asymmetric showing abrupt changes. Furthermore, one observes large deviations between the maxima and also large deviations between the minima in the $\left[f_2 - t \right]$ plot, shown in Fig. 4d. These characteristics suggest chaotic motion. The above results, strongly suggest that the $f_2$-indicator gives reliable results for borderline chaotic orbits, in the global potential. The integration time of the orbit and the $S(c)$ spectrum shown in Figs. 4a and 4c is 200 time units and $2 \times 10^4$ respectively.

The next step is to present results for the 2D local potential (3.3). Figure 5a-d is similar to Fig. 3a-d but for an orbit in potential (3.3), with initial conditions: $x_0=0, p_{x0}=0.3$. The value of $p_{y0}$ is always found using the Hamiltonian (3.4). The values of all other parameters are as in Fig. 2a. All three indicators show that the orbit is regular. The integration time of the orbit and the $S(c)$ spectrum shown in Figs. 5a and 5c are same as in Figs. 3a and 3d respectively. On the other hand, the results given by three indicators in Figure 6a-d, suggest that the corresponding orbit is chaotic. The initial conditions of the above orbit are: $x_0=0.47, p_{x0}=0$, while the values of all other parameters are as in Fig. 2b. As this orbit is a chaotic orbit, we conclude, that the $f_2$-indicator gives reliable results for chaotic orbits both in the global and local potentials. The integration time of the orbit and the $S(c)$ spectrum is same as in Figs. 4a and 4c.

The above procedure was done for a large number of orbits (about 1000) in the global and local potentials. The outcomes strongly suggest, that the $f_2$-indicator can be safely used in order to detect the regular or chaotic character of a 2D orbit. In all cases, the results given by the $f_2$-indicator were found to coincide with the outcomes given by the L.C.E and the $S(c)$ spectrum. Here we must note, that the method of the $f_2$-indicator is faster than the other two methods, as we need 2000 - 5000 time units in order to obtain reliable results, while for the $S(c)$ dynamical spectrum we need an average time of 5000 - 20000 time units, in order to obtain reliable results. Furthermore, the $S(c)$ spectrum needs the calculation of the phase plane produced by the orbit, while the $f_2$-indicator needs only the orbit. On the other hand, it is beyond any doubt, that the $f_2$-indicator is much more faster than the L.C.E, which needs time periods of order of $10^5$ time units, in order to give reliable results.

\section{The character of motion in the 3D models}

In this Section we shall investigate the regular or chaotic nature of orbits in the 3D potentials. We shall start our investigation from the global potential (2.1). All orbits were started with initial conditions $\left(x_0,p_{x0},z_0\right), y_0 = p_{z0}=0$, where $\left(x_0,p_{x0}\right)$, is a point on the phase planes of the 2D system. This point lies inside the limiting curve
\begin{equation}
\frac{1}{2}p_x^2+V_G(x)=E_2 \ \ \ ,
\end{equation}
which is the curve containing all the invariant curves of the 2D system. We use $E=E_2$, where $E$ is the numerical value of the Hamiltonian for the global 3D potential and the value of $p_{y0}$, for all orbits, is obtained from the Hamiltonian (2.4). The numerical calculations suggest that orbits with initial conditions $\left(x_0,p_{x0},z_0\right), y_0 = p_{z0}=0$, such as $\left(x_0,p_{x0}\right)$, is a point in the chaotic regions of Figure 1a-b, for all permissible values of $z_0$, produce chaotic orbits. On the other hand, orbits with initial conditions $\left(x_0,p_{x0},z_0\right), y_0 = p_{z0}=0$, such as $\left(x_0,p_{x0}\right)$, is a point in the regular regions of Figure 1a-b, behave in quite a different way. Orbits with initial conditions, such as $\left(x_0,p_{x0}\right)$, is a point in the regular region around the two elliptic periodic points, on the $x$ axis, are regular for values of $z_0 \lesssim 2.7$, while for larger values of $z_0$ they become chaotic. On the other hand, orbits with initial conditions in all other regular regions of Figs. 1a-b, give regular orbits, for $z_0 \lesssim 0.46$, while they become chaotic when $z_0>0.46$.

Figure 7a-d shows the orbit (a), the L.C.E (b), the $S(c)$ dynamical spectrum (c) and the $f_3$-indicator (d), for an orbit in the 3D global potential. The initial conditions are: $x_0=-1.5, p_{x0}=15, z_0=0.1$, while the values of all other parameters are as in Fig. 1b. It is important to note that, all three dynamical indicators show that the nature of the orbit is chaotic. The integration time for all 3D orbits is 200 time units, while for the $S(c)$ spectrum is $2 \times 10^4$ time units. Figure 8a-d is similar to Fig. 7a-d, for an orbit in the same potential with initial conditions: $x_0=-7.4, p_{x0}=0, z_0=0.1$. The values of all other parameters are as in Fig. 1b. Here we see a quasi periodic orbit, characteristic of the 2:3 resonance. As expected, once more, all three indicators coincide, showing that the motion is regular. Figure 9a shows an orbit with all initial conditions and parameters same as those used in Fig. 8a, but for $z_0=0.6$. Note that, here the character of the orbit has changed from regular to chaotic. It is obvious, that the chaotic nature of this orbit is suggested by all dynamical parameters.

%%%%%%%%%%%%%%%%%

Before closing this Section, we shall present some numerical results for the 3D local potential. Fig. 10a-d shows a chaotic orbit (a) and the corresponding indicators (b-d). The initial conditions are: $x_0=0.48, p_{x0}=0, z_0=0.1$. The values of all other parameters are as in Fig. 2b. Obviously, the motion is chaotic. In Figure 11 a-d we can see a 3D regular orbit with initial conditions: $x_0=0.2, p_{x0}=0, z_0=0.1$. The values of all other parameters are as in Fig. 2a. In this case, the motion is regular. Finally, the orbit presented in Fig. 12a has initial conditions and all other parameters same as those used in Fig. 11a, but $z_0=0.6$. Here the motion becomes chaotic. Generally speaking, the motion in the 3D local potential for orbits with initial conditions $\left(x_0,p_{x0},z_0\right), y_0 = p_{z0}=0$, such as $\left(x_0,p_{x0}\right)$, is a point in the regular regions of Figure 2a give regular orbits  when $z_0 \leq 0.26$, while for larger values of $z_0$ the motion becomes chaotic. On the other hand, orbits with initial conditions $\left(x_0,p_{x0},z_0\right), y_0 = p_{z0}=0$, such as $\left(x_0,p_{x0}\right)$, is a point in the regular regions of Figure 2b are chaotic, when $z_0 \gtrsim 0.17$, while for $z_0 < 0.17$ they remain regular.

The above procedure was done for a large number of orbits, both in the global and local 3D potentials. For a grid of $100 \times 100$ equally spaced initial conditions $(x_0, p_{x0})$ on the $(x,p_x), y=0, p_y>0$ phase plane and for all permissible values of $z_0$, we need about 4h of CPU time on a Pentium IV 2GHz PC. The results strongly suggest, that the $f_3$-indicator can be safely applied in order to distinguish between ordered and chaotic motion in 3D Hamiltonian systems. In all cases, the outcomes derived by the $f_3$-indicator, were found to coincide with the results given by the L.C.E and the $S(c)$ dynamical spectrum.

\section{Discussion and conclusions}

In the present article, we have studied the regular or chaotic character of orbits in two 3D galactic autonomous Hamiltonian systems. The global potential (2.1) and the local potential (2.2). The novelty of this research is that in this paper, we have introduced and used a new dynamical parameter, the $f$-indicator, in order to detect the regular or chaotic nature of motion, in different kinds of Hamiltonian systems. In general, the nature of an orbit in 2D or 3D dynamical systems, can be identified using the $f_2$ or $f_3$ indicator respectively. The character of motion can be visualized by the profile of the $\left[f - t \right]$ plot. If the motion is ordered, it produces a quasi periodic profile, with almost symmetrical peaks. On the contrary, in the case of chaotic motion, the profile produced by the $f$-indicator, is highly asymmetric with large deviations between the maxima and also significant deviations between the minima in the $\left[f - t \right]$ plot. The importance of this new dynamical indicator, is that we can obtain fast and reliable results regarding the nature of motion, especially in Hamiltonian systems with three degrees of freedom, in which the structure of the phase space is complicated and therefore, hard to interpret.

In both galactic systems (2.1) and (2.2) we have started our investigation from the 2D potentials. It was observed that the global 2D potential displays considerable chaotic regions together with areas of regular motion. Responsible for the chaotic motion is the asymmetric term $-\lambda x^3$. In the local potential the chaotic regions are small when $\gamma =-0.2$, while for $\gamma=-0.8$ a large chaotic sea is observed and the regular regions are small. Our numerical experiments have shown that the $f_2$-indicator gives fast and reliable results regarding the nature of motion in both 2D galactic potentials. Comparison with the L.C.E and the $S(c)$ dynamical spectrum, justifies the reliability of the $f_2$-indicator.

The $f_3$-indicator is also a useful tool for distinguishing between order and chaos in 3D dynamical systems. Comparison with the L.C.E and the $S(c)$ spectrum, shows that all three dynamical parameters give similar results. This shows that the new indicator can be trusted, in order to distinguish regular from chaotic orbits. What is interesting is that the L.C.E needs time periods of order of $10^5$ time units and the $S(c)$ periods of order $10^4$ time units, while the $f$-indicator give reliable results for time periods of between 2000-5000 time units. Furthermore, the $f$-indicator needs comparable time of numerical integration as the SALI, FLI and GALI methods, while it is much more faster than the dynamical spectra of stretching numbers.

Before closing, the author would like to make clear that the $f$-indicator, as defined in Section 2, was based on the energies along the $x, y$ and $z$ axis ignoring the coupling between the three variables $x, y$ and $z$. But all calculations were made in the 3D potentials (2.1),(2.2) and the 2D potentials (3.1),(3.3), where the coupling does exist. It is evident, that the complicated mechanism of this coupling is responsible for the regular or chaotic nature of orbits, shown very clearly by the $f$-indicator.

It is our hope, to be able to apply in the future, the new $f$-indicator in other kinds of interesting potentials and in more complicated dynamical systems, such as those describing binary stellar systems (see Caranicolas and Innanen, 2009). With additional theoretical work, we will investigate if it always distinguishes correctly between ordered and chaotic motion, in Hamiltonian systems, especially in dynamical systems with three degrees of freedom. Therefore, one may conclude, that the $f$-indicator is a very fast and reliable tool, which can be applied not only in galactic dynamics, but in non-linear dynamics in general.

\section*{Acknowledgements}
{The author would like to thank the anonymous referee for the careful reading of this manuscript and for his very useful suggestions and comments, which improved the quality of the present article.}

\label{pagefin}
\end{document}